\documentclass[a4paper,fleqn,usenatbib]{mnras}
\usepackage{newtxtext,newtxmath}

\usepackage[T1]{fontenc}
\usepackage{ae,aecompl}

\usepackage{caption}
\usepackage{subcaption}
\usepackage{pdflscape}
\usepackage{comment}

\usepackage{graphicx}	
\usepackage{amsmath}	
\usepackage{siunitx}
\usepackage{float}
\usepackage[dvipsnames]{xcolor}
\usepackage{multirow}

\title[Halo shapes in self-interacting dark matter]{Constraining SIDM with halo shapes: revisited predictions from realistic simulations of early-type galaxies}

\author[Despali, et al. 2022]{Giulia Despali$^{1}$\thanks{E-mail:gdespali@uni-heidelberg.de }, 
Levi G. Walls$^{2,3}$, Simona Vegetti$^{3}$, Martin Sparre$^{4,5}$, \newauthor Mark Vogelsberger$^{6}$, Jes\'{u}s Zavala$^{7}$
\\\\
$^{1}$ Institut f\"{u}r Theoretische Astrophysik, Zentrum f\"{u}r Astronomie, Heidelberg Universit\"{a}t, Albert-Ueberle-Str. 2, 69120, Heidelberg, Germany\\
$^{2}$ Department of Astronomy, University of Michigan, 323 West Hall, 1085 S. University, Ann Arbor, MI 48109 USA
\\
$^{3}$ Max Planck Institute for Astrophysics, Karl-Schwarzschild-Strasse 1, 85748 Garching bei M\"{u}nchen, Germany\\
$^{4}$ Institut f\"{u}r Physik und Astronomie, Universit\"{a}t Potsdam, Karl-Liebknecht-Str. 24/25, D-14476 Golm, Germany\\
$^{5}$ Leibniz-Institut für Astrophysik Potsdam (AIP), An der Sternwarte 16, 14482 Potsdam, Germany\\
$^{6}$ Department of Physics, Kavli Institute for Astrophysics and Space Research, Massachusetts Institute of Technology, Cambridge, MA 02139, USA\\
$^{7}$ Center for Astrophysics and Cosmology, Science Institute, University of Iceland, Dunhagi 5, 107 Reykjavík, Iceland
}

\date{Accepted 2022 September 1. Received 2022 September 1; in original form 2022 April 25}

\pubyear{2022}

\begin{document}
\label{firstpage}
\pagerange{\pageref{firstpage}--\pageref{lastpage}}
\maketitle
\allowdisplaybreaks
\begin{abstract}
We study the effect of self-interacting dark matter (SIDM) and baryons on the shape of early-type galaxies (ETGs) and their dark matter haloes, comparing them to the predictions of the cold dark matter (CDM) scenario. We use five hydrodynamical zoom-in simulations of haloes hosting ETGs ($M_{\rm vir}\sim 10^{13}M_{\odot}$ and $M_{*}\sim10^{11}M_{\odot}$), simulated in CDM and a SIDM model with constant cross-section of $\sigma_T/m_\chi = 1\ \mathrm{cm}^2 \mathrm{g}^{-1}$. We measure the three-dimensional and projected shapes of the dark matter haloes and their baryonic content using the inertia tensor and compare our measurements to the results of three \emph{HST} samples of gravitational lenses and \emph{Chandra} and \emph{XMM-Newton} X-ray observations. We find that the inclusion of baryons greatly reduces the differences between CDM and a SIDM, together with the ability to draw constraints based on shapes. Lensing measurements reject the predictions of CDM dark-matter-only simulations and prefer one of the hydro scenarios. When we consider the total sample of lenses, observational data prefer the CDM hydro scenario. The shapes of the X-ray emitting gas are compatible with observational results in both hydro runs, with CDM predicting higher elongations only in the very centre. Contrary to previous claims at the scale of elliptical galaxies, we conclude that both CDM and our SIDM model can still explain observed shapes once we include baryons in the simulations. Our results demonstrate that this is essential to derive realistic constraints and that new simulations are needed to confirm and extend our findings.

\end{abstract}

\begin{keywords}
cosmology: dark matter - methods: numerical - galaxies:halos -  galaxies: elliptical and lenticular, cD - X-rays: galaxies - gravitational lensing: strong
\end{keywords}



\section{INTRODUCTION} \label{sec:intro}
The foundation on which our current understanding of structure formation is based is the $\Lambda$CDM model, which describes the Universe as filled with cold dark matter (CDM) that is collisionless and interacts only via the gravitational force. The CDM paradigm is widely successful in explaining many aspects of galaxy formation and evolution \citep{springel05b,schaye15,vogel14,pillepich18,vogel20}. 
However, the corresponding particles (e.g. WIMPs) required by the CDM model have not yet been found with direct or indirect detection experiments \citep{2018RPPh...81f6201R}. Consequently, other models have been proposed as possible alternatives to CDM, also justified by the well-known tensions between $\Lambda$CDM predictions and observations on cosmological \citep{verde19} or small \citep{bullock17} scales. 

Self-interacting dark matter (SIDM) postulates that dark matter particles are not collisionless but can have strong interactions and exchange energy and momentum. SIDM was originally invoked to address the discrepancies between the observed properties of dwarf galaxies and the predictions from (DM-only) CDM simulations and has the potential to solve many of the small-scale CDM problems \citep[for a review see][]{tulin18}. The term SIDM refers to a variety of models that can include elastic  or inelastic scattering, a constant or velocity-dependent  interaction cross-section  \citep{vogel12,vogel16,vogel19,Despali:2019,sameie18,sameie20,rocha13,robertson18,robertson21,lovell19,kaplinghat14,kaplinghat19,kaplinghat20,correa21}. 

The most important signature of SIDM across all models is the creation of a central core in the density profile of haloes, due to the momentum and energy exchanges between the particles, with an extent that depends on the  self-interaction cross-section and the details of the model \citep{colin02,rocha13,vogel12,vogel14b,vogel16}. This process is efficient in the centre of haloes where the density is high and leads to modifications of the density cusp predicted by the standard NFW profile \citep{navarro97}. In the outskirts, where the density drops, the effects are negligible, and the density profile follows the NFW prediction. 
However, this simple picture is modified once the effect of baryons is considered: baryonic matter dominates the central region of haloes and, as demonstrated by CDM hydrodynamical simulations, can significantly alter the halo properties predicted by dark-matter-only simulations. In CDM, the gradual growth of the baryonic potential has the effect of adiabatically contracting the dark matter halo and increasing its concentration \citep{blumenthal86,gnedin04,schaller15,lovell19}, while rapid events such as feedback and supernova explosions can lower the central density \citep{mashchenko06,pontzen10,burger21,2016MNRAS.459.2573R,2016MNRAS.456.3542T,2015MNRAS.454.2092O}. The inclusion of baryons can also alleviate some of the tensions between CDM and observations, such as the classic missing satellites and too-big-to-fail problems \citep{zolotov12,kim18,garrison19} and create cored profiles \citep{2019MNRAS.488.2387B}. In SIDM, the role of the dark matter and baryonic potentials depends on the self-interaction cross-section and on the timescale of SIDM interactions compared to the time evolution of the dark matter density profile. Recent works have started to explore the interplay between SIDM and baryons, using hydrodynamical simulations at galaxy \citep{sameie18,Despali:2019} and galaxy cluster \citep{robertson18,robertson21,shen22} scales. The inclusion of baryons generally reduces the differences between CDM and SIDM predictions with respect to dark-matter-only simulations: the baryonic potential of the central galaxy partially counteracts the flattening due to self-interactions. 

In this work, we focus on another halo property affected by self-interactions: halo shapes. In dark-matter-only simulations, the self-interactions between particles produce rounder central shapes than in the CDM scenario \citep{dave01,Peter:2013II,brinckmann18}. 
Previous works have used gravitational lensing \citep{miralda02} and X-ray \citep{buote02,mcdaniel21} observations to estimate the ellipticity (or triaxiality) of elliptical galaxies or clusters and set some of the strongest constraints in the literature on the allowed range of self-interaction cross-sections. Other recent works focus instead on similar constraints at the scale of galaxy clusters \citep{robertson18, shen22,andrade22, eckert22} or on the intrinsic alignments of galaxy shapes \citep{harvey21}. The advantage of considering elliptical galaxies is twofold: early-type galaxies are common lens galaxies at high-redshift \citep{auger10b} and are well described by a smooth mass distribution \citep{enzi20}; moreover, in relaxed elliptical galaxies, it is reasonable to assume that the X-ray emitting interstellar gas is in hydrostatic equilibrium with the halo potential. Under this condition, the shape of the X-ray emission can be used to derive the overall shape of the gravitational potential. Applying this method to the elliptical galaxy NGC 720, \citet{buote02} determined a halo ellipticity of $\epsilon\simeq0.35-0.4$ from \emph{Chandra} X-ray data. Comparison to theoretical predictions for NFW and Hernquist profiles led to the conclusion that this value was consistent with an interaction cross-section of $\sigma_T/m_\chi\sim0.1$ cm$^{2}$/g as well as with CDM, while higher cross-sections were disfavoured. A similar conclusion was reached by \citet{Peter:2013II}, who compared these observational data to dark-matter-only simulations in CDM and SIDM.  More recently, \citet{mcdaniel21} analysed \emph{Chandra} and \emph{XMM-Newton} observations of a sample of eleven elliptical galaxies and compared their X-ray surface brightness profile to the predicted emission generated by dark matter haloes with varying density profiles and triaxiality. Their findings led to milder constraints with respect to previous studies and showed that a self-interaction at the scale of $\sigma_T/m_{\chi}\sim1 $cm$^{2}$/g would still be consistent with their observations (as well as the CDM scenario).

One significant limitation of most previous works at the scale of massive galaxies is that the simulations used for the comparison with observations did not include baryonic physics. It is well known that the inclusion of baryons can significantly alter the central shapes of CDM haloes \citep{chua19}, leading to rounder configurations with respect to CDM dark-matter-only runs. In practice, baryons and self-interactions both affect halo shapes in a similar manner and thus their combination can alter the distribution of predicted shapes \citep{chua21}, leading to different constraints.
The aim of this work is to improve the computational predictions at the scale of elliptical galaxies ($M_{\rm vir}\sim 10^{13}M_{\odot}$ and $M_{*}\sim10^{11}M_{\odot}$), by analysing the hydrodynamical zoom-in simulations from  \cite{Despali:2019}. These are the first simulations of early-type galaxies that include both baryonic physics \citep[following the TNG model, see ][]{pillepich18,weinberger18} and self-interacting dark matter (with a constant cross-section  of $\sigma_T/m_{\chi}=1$cm$^{2}$/g). We measure the intrinsic and projected shapes from the mass distribution, and the ellipticity of the simulated X-ray emission from mocks that reproduce \emph{Chandra} and \emph{XMM-Newton} data. We compare our findings in CDM and SIDM with the shapes measured in previous works using gravitational lensing \citep{auger10b,sonnenfeld13,ritondale19a} and X-ray data quality \citep{buote02,mcdaniel21}. In this respect, our work is similar to the recent analysis by \citet{shen22}, who used SIDM simulations with a similar cross-section and including adiabatic gas to compare the shape of the X-ray emission to observations of galaxy clusters.

The paper is organised as follows. In Section \ref{sec:Simulations} we describe the simulations and Section \ref{sec:Methods} sets out our analysis process. In Section \ref{sec:Shapes} we present the measurements of three-dimensional and projected shapes from the simulations and compare them with observational data; in Section \ref{sec:Xray} we analyse the X-ray mock images and compare them to real observations by \emph{Chandra} and \emph{XMM-Newton}. Finally, we discuss our results and draw our conclusions in Section \ref{sec:summary}. In the Appendix, we discuss technical aspects in more detail.

\section{Simulations}	
\label{sec:Simulations}

\begin{figure*}
    \centering
	\includegraphics[width=0.45\hsize]{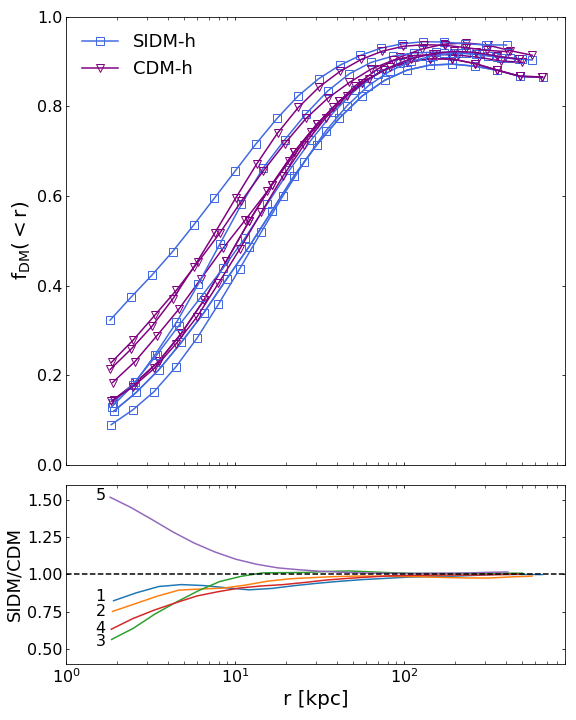}
	\includegraphics[width=0.45\hsize]{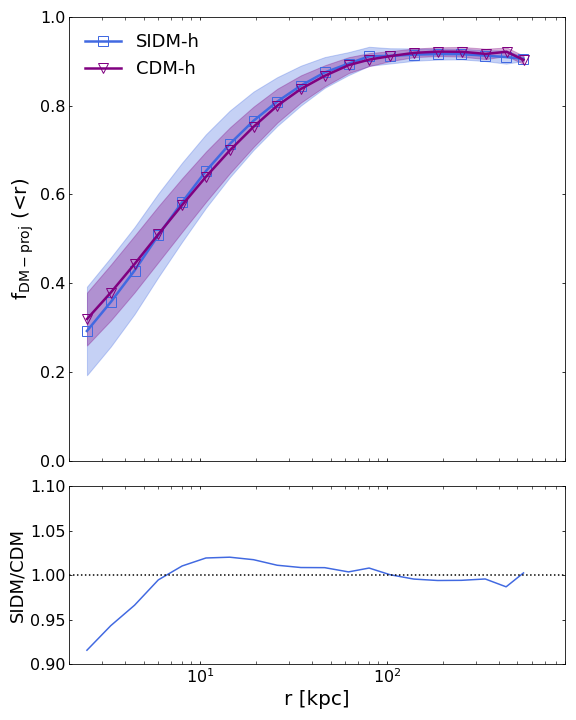}
	\caption{Fraction in dark matter as a function of radius in the CDM-h (purple, triangles) and SIDM-h (blue, squares) runs that include baryons, calculated in logarithmically spaced bins in the range $[0.01,1]\times R_{\rm vir}$. The left panel shows the dark matter fraction for each halo in both scenarios, whereas the right panel shows the average projected fraction over 90 viewing angles for each halo. Points and solid lines show the average fraction over the whole sample of haloes and projections, together with the standard deviation (shaded region). In the bottom panels, we show the ratio between the SIDM-h and CDM-h cases.}
	\label{fig:frac}
\end{figure*}

\begin{table*} 
\centering

\caption{Summary of halo properties in the CDM and SIDM runs: SUBFIND IDs in the parent Illustris simulations, total halo mass $M_\text{vir}$, stellar mass of the central galaxy $M_{*}$ for the two hydro runs and virial radius expressed in \emph{comoving} kpc. We refer the reader to \citet{Despali:2019} for more details on each system. throughout the paper, CDM-d and SIDM-d are the dark-matter-only runs, while CDM-h and SIDM-h their counterparts including baryonic physics. \label{tab:HaloID}}
\begin{tabular}{|c|cccc|cccc|cc|} 
\hline 
halo & \multicolumn{4}{c|}{M$_\text{vir}[10^{13}M_{\odot}]$} & \multicolumn{4}{c|}{R$_\text{vir}$ [kpc]} &\multicolumn{2}{c|}{M$_{*}[10^{11}M_{\odot}]$}\\\ 
ID &  CDM-d & SIDM-d & CDM-h & SIDM-h& CDM-d & SIDM-d & CDM-h & SIDM-h & CDM-h & SIDM-h \\
\hline  
1 & 2.66 & 2.67 &3.73 & 3.76 & 940 & 941 & 937 & 938  & 4.08  & 4.50\\
2 & 2.13 & 1.81 & 2.45 & 2.46 & 872 & 835 & 815 & 813 & 2.71  & 3.02 \\
3 & 1.24 & 1.24 & 1.64 & 1.65  & 730 & 728 & 714 & 713 & 2.16 & 1.75\\
4 & 1.10 & 1.10 & 1.4 & 1.41  & 698 & 700 & 678 & 676 & 3.03 & 3.04\\
5 & 0.71 & 0.74 & 0.91 & 0.86  & 605 & 613 & 585 & 572 & 1.46 & 1.42\\
\hline

\end{tabular}

\end{table*}

We use the sample of simulated galaxies from \citet{Despali:2019}. These are zoom-in re-simulations of systems chosen from the Illustris simulations \citep{vogel14,Torrey14,Genel14} on the basis of their properties \citep{despali17b} to resemble the lens galaxies from the SLACS survey \citep{bolton06}: haloes that host early-type galaxies (ETGs) at $z\simeq 0.2$ with total mass, stellar mass, stellar effective radius and velocity dispersion consistent with those of SLACS lenses. Among this population, \citet{Despali:2019} have re-simulated a subsample of nine galaxies with the zoom simulation method described in \citet{2016MNRAS.462.2418S}: the dark matter mass resolution is $4.4\times 10^6 M_{\odot}$, while the baryonic resolution is $9.1\times 10^5 M_{\odot}$. Each galaxy was re-simulated both in the standard CDM scenario and in a SIDM model with a constant cross-section of $\sigma_\text{T}/m_\chi= 1\; \text{cm}^{2}\text{g}^{-1}$, using an extended version of the AREPO code that includes both elastic and multi-state inelastic SIDM~\citep{vogel12,vogel16,vogel18a}. The IllustrisTNG physics model \citep{pillepich18,weinberger18} was used to describe the evolution of baryonic physics, including AGN feedback \citep{2017MNRAS.465.3291W} and magneto-hydrodynamics \citep{2013MNRAS.432..176P,2020MNRAS.499.4261S}. Similarly to the Illustris model \citep{vogel14}, the TNG physics was calibrated to reproduce observables related to the stellar content of galaxies, and the subgrid model contains prescriptions for processes at scales below the resolution \citep[see ][ for more details]{pillepich18}. Haloes and subhaloes have been identified with the Friends-of-friends (FoF) method by SUBFIND \citep{springel05b}; a list of the main halo and galaxy properties can be found in Table \ref{tab:HaloID}. We define the virial mass $M_{\text{vir}}$ as the mass within the virial radius $R_{\text{ vir}}$ that encloses  a  virial overdensity $\Delta_{\rm vir}$, calculated following \citet{bryan98}. 

For this work, we also ran the SIDM dark-matter-only version of the each simulation, while we use the original Illustris1-Dark simulation as the CDM dark-matter-only counterpart. In this case, the dark matter mass resolution is $7\times10^{6}M_{\odot}$. For simplicity, throughout the text, we refer to the dark-matter-only simulations as CDM-d and SIDM-d, while to their counterparts that include baryonic physics as CDM-h and SIDM-h. 

Some of the galaxies that were identified as ETGs in the original Illustris-1 run develop disks in our zoom-in simulations, due to the interplay between SIDM and baryons and the differences between the Illustris and the IllustrisTNG model \citep[see][for more details]{Despali:2019}. In this work, we focus on five galaxies which maintain ETG-like morphologies in both scenarios for an effective comparison with observational results based on ellipticals and ETGs. We point out that we have not re-calibrated the galaxy formation model in the case of SIDM: the properties of the five galaxies used here well reproduce those of observed ellipticals in both dark matter scenarios (see Table \ref{tab:HaloID}), but we cannot predict if the properties of the galaxy population as a whole (for example the disk fraction) would be the same in a cosmological box. We prefer not to speculate here on the potential differences beyond what had been already found by previous works: for example, we know from \citet{robertson21,eckert22} that the profiles of galaxy clusters are modified by a SIDM model similar to the one considered here, while galaxies with mass $M_{\text{vir}}\sim10^{13}M_{\odot}$ are the most similar in the two models, because baryons dominate their inner halo profiles. In this work we focus on relative differences between the two dark matter models, and we leave other considerations for future work. 

\citet{Despali:2019} studied the density profile and formation history of these simulated haloes and classified them according to the change in central density from CDM to SIDM. 
Haloes were classified into two main categories: $(i)$ "CORED" systems, i.e. with lower central density in SIDM with respect to CDM, associated with a low formation redshift and $(ii)$ "CUSPY" system with, conversely, higher formation redshift and higher central density in SIDM. They found that these properties correspond to a difference in the halo mass accretion history and formation redshift: CORED (CUSPY) systems had lower (higher) formation redshift. Amongst the haloes considered in this work, systems 1-4 belong to the first category (CORED), while halo 5 to the second (CUSPY). The properties of halo 5 are similar to those of the other CUSPY haloes that were used in \citep{Despali:2019}, but not included in this work - and we refer the reader to that work for more details on the mass accretion history. In the left panel of Figure \ref{fig:frac}, we show the dark matter fraction as a function of radius for the CDM-h (purple triangles) and SIDM-h (blue squares) runs; halo 5 is the only case of a higher central dark matter fraction in SIDM, given that it belongs to the CUSPY group. When the dark matter fraction is measured in projection (right panel), the differences between the two dark matter models are greatly reduced with respect to the 3D case. We expect that the effect of baryons on the halo shapes will be especially relevant within $\simeq$30 kpc from the centre, both in 3D and in projection.

The zoom-in simulations used in this work reach $z=0.2$, corresponding to the characteristic redshift of the SLACS lenses \citep{bolton06}. While this is ideal in the case of a comparison with gravitational lenses (i.e. high redshift galaxies), it might introduce a bias in the comparison to X-ray observations of low-$z$ elliptical galaxies. As shown in \citet{despali14}, the average halo ellipticity at fixed mass increases with redshift, but it varies by only $\simeq$10 per cent between $z=0$ and $z=0.5$. Throughout this work, we focus on the one-to-one comparison between our simulated samples in different scenarios, and thus we think that the effect introduced by SIDM and baryons with respect to CDM would be the same or very similar at a different redshift, even if possibly rescaled in magnitude.

\section{Methods}	
\label{sec:Methods}

\begin{table*}
    \centering
    \caption{Summary of the shape definitions used in this work.}
    \begin{tabular}{|c|c|c|c|}
    \hline
    \textbf{Source} & \textbf{Dimensions} & \textbf{Method} & \textbf{Parameters} \\
    \hline
      & 3D & inertia tensor &  axes: $a\geq b\geq c$ \\
     simulations & triaxial ellipsoid & on mass distribution &axis ratios: $s=c/a$, $q=b/a$ \\
     & &  & triaxiality: $T=(1-q^{2})/(1-s^{2})$\\
     \hline
     simulations & 2D & inertia tensor & axes: $a\geq b$\\
      & projected ellipse & using projected mass & ellipticity: $e=1-b/a$  \\
     \hline
     & 3D & inertia tensor & approximation from 3D ellipsoid\\
     simulations & spheroid & on mass distribution & axes:  $a\geq b\geq c$\\
     & with $a=b$ or $b=c$ &  &$\epsilon=1-\sqrt{bc}/a$ if prolate\\
     & & &$\epsilon=1-c/\sqrt{ab}$ if oblate\\
     \hline
      X-ray observations & 2D & inertia tensor& ellipticity $e_{X}=1-b/a$\\
    and mocks &  &  on image pixels & \\
     \hline
     lensing & 2D & elongation of the  & elongation $q=b/a$\\
     observations & SIE+shear & Singular Isothermal Ellipsoid &ellipticity $e=1-q$\\
     \hline
    \end{tabular}
    \label{tab:shape}
\end{table*}

\begin{figure*}
 
	\includegraphics[width=0.9\textwidth]{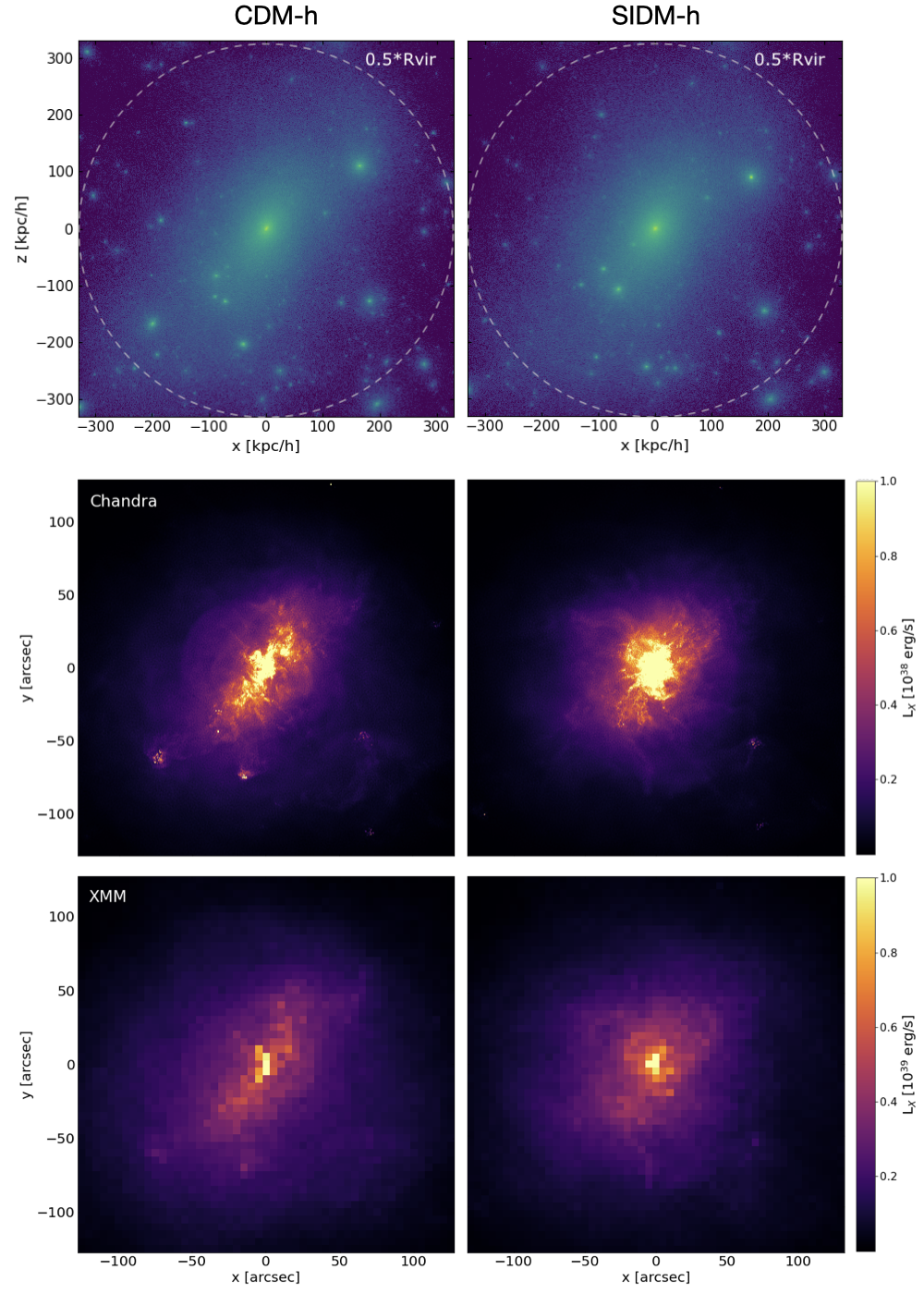}
	
     \caption{Example maps for one system (halo 1) in our sample. The two top panel show the dark matter distribution in the CDM-h and SIDM-h runs. The other four panels show instead the mock X-ray images created using the pipeline described in \citep{Barnes2020}. These are soft band X-ray energies between 0.5 and 2 keV for \emph{Chandra} and \emph{XMM-Newton}, with a resolution of 0.496" and 5" (middle and bottom panels, respectively) and one the same physical scale of the top panels. The images were created with our CDM-h (left) and SIDM-h (right) runs. Note that the Chandra and XMM maps have different limits, respectively in units of $10^{38}$ and $10^{39}$ erg/s.}
    \label{fig:MockXray}
\end{figure*}
\begin{figure*}
 
	\includegraphics[width=0.9\textwidth]{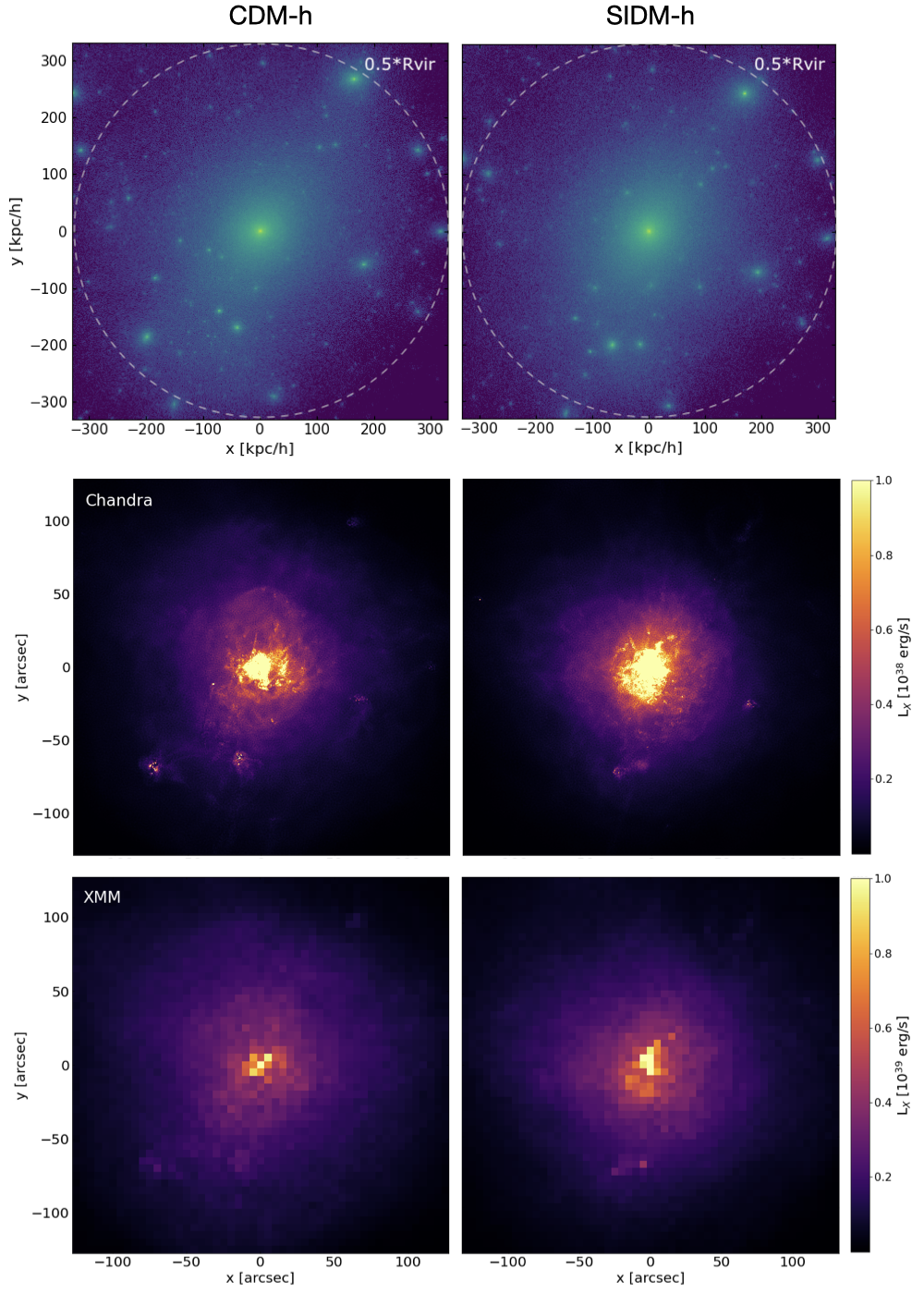}
	
     \caption{Same as Figure \ref{fig:MockXray}, but here the same halo is viewed from a different projection.}
    \label{fig:MockXray2}
\end{figure*}

The goal of this paper is to quantify the effect that SIDM and baryons have on the shapes of ETGs and whether or not CDM and  can be distinguished from a SIDM model with $\sigma_T/m_{\chi}=1$ cm$^{2}$/g on this basis using observations.
In this section we introduce our definitions of halo shapes (see also Table \ref{tab:shape}) and we describe our procedure to create mock X-ray observations from the simulations. Finally, we summarise the observational quantities that we are trying to reproduce and discuss how we compare them to their simulated counterparts, together with the potential limitations. Figures \ref{fig:MockXray} and \ref{fig:MockXray2} show examples of the projected dark matter distribution (top panels) and X-ray emission (middle and bottom panels) for one halo and two different projections.

\subsection{Halo shapes from the inertia tensor}

Previous works demonstrated that dark matter haloes are well described by triaxial ellipsoids, which approximate the radial matter distribution better than spherical symmetry  \citep{allgood06,Despali:2013,Peter:2013II,despali17,chua19}. Similarly to previous works, here we use an iterative method to calculate the halo shapes via the (weighted) inertia tensor, defined as:
\begin{equation}
    I_{\alpha \beta} = \left. \frac{1}{N} \sum_{i=1}^{N} \frac{x_{i,\alpha} x_{i,\beta}m_{i}}{r_{i}^{2}}\right/\sum_{i=1}^{N}m_{i}.
\end{equation}
Here $\mathbf{x}_{i}$ are the position components of the $i$-th particle from the centre of mass, $\alpha$ and $\beta$ are the tensor indices, $m_{i}$ is the particle mass and $r_{i}$ is the distance of the $i$-th particle from the centre of mass of the halo. We evaluate shapes in shells of increasing radius, with logarithmically-spaced radial bins within [0.01,1]$\times R_{\rm vir}$. At each radius, we start from a spherical volume and we evaluate the inertia tensor; in order to avoid perturbations due to substructures, we select only the particles belonging to the main halo (i.e. the first subhalo identified by SUBFIND). The squared roots of the three eigenvalues of the inertia tensor define the three axes of the best-fitting ellipsoid ($a\geq b\geq c$). We define the axis ratios $s=c/a$ and $q=b/a$ and the triaxiality parameter $T=(1-q^{2})/(1-s^{2})$, that measures how prolate ($T=1$) or oblate ($T=0$) the halo is. The eigenvectors determine the spatial orientation of the ellipsoid. 
We then refine the shape iteratively by deforming the considered region into a triaxial ellipsoid: we use the eigenvectors to rotate the particle distribution and align it with the principal axes, then we select the particles contained in an ellipsoid enclosing the same volume of the original sphere, using the ellipsoidal distance defined as
\begin{equation}
    r_{i}^{2} = dx_{i}^{2}/a^{2} +  dy_{i}^{2}/b^{2} + dz_{i}^{2}/c^{2}. \label{eq_rell}
\end{equation}
The inertia tensor is re-evaluated on the new set of particles, and this procedure is repeated until the axis ratios converge at the one per cent level. At each radius, the shape is calculated independently and the axis ratios and orientation of the ellipsoid are free to vary. In each case, we require a minimum of 1000 particles to guarantee a reliable estimate.

We also measure projected shapes following the same procedure on two projected coordinates: in this case, we calculate the best-fit ellipse, described by two axes ($a\geq b$) and the ellipticity $e=1-b/a$. As described in more detail in the next Section, we calculate projected shapes along multiple lines of sight: each time, we rotate the halo by randomly choosing three Euler angles ($\phi,\theta,\psi$), where $\phi$ and $\psi$ are sampled from a uniform distribution between 0 and $2\pi$ and $\theta$ between 0 and $\pi$. For each rotated configuration, we then project the mass distribution along the three principal directions and measure the projected shape.

\subsection{X-ray maps} \label{sec:Methods2}

We use the pipeline from \citet{Barnes2020} to create mock X-ray images from our CDM-h and SIDM-h simulations, generating the X-ray emission directly from the properties of the gas. We generate mocks  corresponding to the properties of \emph{Chandra} and \emph{XMM-Newton} observations in the soft X-ray energy band 0.5-2 keV. 

 We begin by generating an X-ray spectrum for every gas particle/cell within the FoF group via a lookup table of spectral templates. We generate the table using the Astrophysical Plasma Emission Code \citep[APEC,][]{smith01}. For every particle/cell, we compute a total spectrum using its temperature, density and metal abundance by summing over the chemical elements tracked by the simulations. The spectrum for every particle is then projected down the relevant axis and smoothed onto a square grid. In each projection, the centre of the images is taken to be the projection of the centre of the gravitational potential. The pixel size is set to 0.496" arcsec for \emph{Chandra} and 5" for \emph{XMM-Newton}; following common choices from the literature we convolve the maps with a Gaussian PSF, where the FWHM is assumed to equal the pixel size in both cases. We assume that the mock images do not contain noise or background.

We show an examples of mock images for halo 1, viewed in two different projections in Figure \ref{fig:MockXray} and \ref{fig:MockXray2}. The middle and bottom panels show the mock X-ray emission as would be observed by \emph{Chandra} and \emph{XMM-Newton} in the soft-bands 0.5-2 keV, from the CDM-h (left) and SIDM-h (right) runs. Comparing the two figures, we can appreciate the effect of looking at the same halo from a different viewing angle: in Figure \ref{fig:MockXray} the CDM-h and SIDM-h cases appear very different, whereas the shapes of the X-ray emission are much closer in Figure \ref{fig:MockXray2}.

We  measure the shapes of the X-ray emission directly from the mock images, following the approach by \citet{buote02}. In practice, we calculate the inertia tensor, but we weight each pixel (instead of each particle) by its flux (instead of mass). We start by measuring the shapes in circular aperture of increasing radius and, at each radius, the shape of the shell is deformed iteratively into an ellipse with the appropriate orientation and ellipticity. Under the assumption of hydrostatic equilibrium (as discussed in more detail in the next Section), these shapes trace the total gravitational potential of the halo.

\subsection{Observed quantities}

Here we describe the observed quantities that we want to reproduce and previous observational works. We discuss how they are compared to the simulations, from the point of view of X-ray observations and gravitational lensing. The shape definitions quoted throughout the paper are summarised in Table \ref{tab:shape}.

\subsubsection*{\textbf{X-ray emission}}

In order to determine the total halo mass through observations of the X-ray emitting gas, one has to rely on the assumption that the gas is in hydrostatic equilibrium and therefore traces the total gravitational potential - a reasonably safe assumption for relaxed elliptical galaxies. Previous works \citep{buote02,mcdaniel21} measured the shape of the X-ray emission from \emph{Chandra} and \emph{XMM-Newton} observations of elliptical galaxies, by fitting the equivalent of the inertia tensor to the surface brightness distribution in the images. We apply the same technique to the mock images generated from simulations (see Sec \ref{sec:Methods2}) and derive ellipticity profiles $e_{X}(r)$. The results of the comparison are presented and discussed in detail in Section \ref{sec:Xray}.

An alternative approach is to find the 3D halo shape that best explains the observed ellipticity profile. However, since projected shapes do not probe the elongation along the line-of-sight, it is not possible to directly measure the three-dimensional shape. Previous works describe the halo as an ellipsoid, but only consider the oblate ($b=c$) and prolate ($b=a$) limit configurations that bracket the possible projected ellipticities of a triaxial ellipsoid: in this way, one defines the projected \emph{spheroidal} ellipticity, but the axes are defined in 3D. \citet{Peter:2013II} converted the triaxial shape measured via the inertia tensor into the \emph{spheroidal} ellipticity in the following way: if the triaxial shape is closer to prolate, then $\epsilon=1-\sqrt{bc}/a$, while if the shape is closer to oblate $\epsilon=1-c/\sqrt{ab}$. These choices set the spheroidal volume equal to the ellipsoidal one. In Section \ref{sec:Shapes}, we use the same approach and calculate $\epsilon$ from our simulations and compare to the results by \citet{buote02,mcdaniel21}.

\subsubsection*{\textbf{Gravitational lensing}} \label{sec:Methods3}
In strong galaxy-galaxy lensing, the light of a background source is deflected and magnified by the lens galaxy. Bright elliptical/early-type galaxies are the typical high-redshift lens galaxies. The entire mass distribution (luminous and dark) contributes to the deflection and can be measured when modelling the lensed images. From the lens modelling, one can reconstruct the two-dimensional lensing potential and the lensing convergence $\kappa$, i.e. the Laplacian of the lensing potential. In practice, the convergence is defined as a dimensionless surface density and so effectively corresponds to a scaled projected mass density, characterising the lens system. It can be written as
\begin{equation}
\kappa(x)= \frac{\Sigma(x)}{\Sigma_\rmn{crit}}, \qquad \rmn{with} \qquad \Sigma_\rmn{crit}= \frac{c^{2}}{4\pi G} \frac{D_\rmn{S}}{D_\rmn{LS}D_\rmn{L}}, \label{eq:lens}
\end{equation}
where $\Sigma_\rmn{crit}$ is the critical surface density and $D_{L}$, $D_{S}$ and $D_{LS}$ stand respectively for the angular diameter distance of the lens, the source and between the lens and the source. The value of the lensing convergence in practice determines by how much the background sources appear magnified on the lens plane. For non-spherical mass distributions, the lensed images of the source are also stretched and distorted along privileged directions by the shear, $\gamma$. 

The shape of the lensing convergence should be in practice equivalent to the shape of the total projected mass distribution calculated via the inertia tensor. One important difference lies in the fact that the density profiles of lenses is commonly modeled as a singular isothermal ellipsoid (SIE), with an elongation $q$, constant with radius. While we do not include assumptions on the density profile in the shape calculation, we indeed impose an elliptical shape. Lensing measurements only probe the mass distribution very close to the galaxy centre, and thus we do not expect a significant variation of the density profile in this range. Moreover, it has been shown \citep{vandeven09} that for a range of (also non-isothermal) stellar and dark matter density profiles, the isothermal approximation works well around the Einstein radius and beyond \citep{gavazzi07}. 

In Section \ref{sec:Shapes}, we compare our measurements with observational results from three samples of strong gravitational lenses observed with HST: the SLACS \citep[][80 lenses]{bolton06,auger10b}, the BELLS-GALLERY \citep[][16 lenses]{ritondale19a} and SL2S \citep[][56 lenses]{sonnenfeld13} lenses, respectively at $z_{l}\sim0.2$, $z_{l}\sim0.5$ and $0.5\leq z_{l} \leq 0.8$. We discuss additional technical details of this comparison in Appendix \ref{sec:App1}.

We take the best lens model from previous works: a SIE (with fixed radial slope equal to -2) was used in the modelling of the SLACS and SL2S samples, while the BELLS sample was modelled with a power-law profile with free slope - the resulting range of radial slopes is $[1.898,2.159]$. In both cases, the elongation $q$ and thus ellipticity $e=1-q$ of the projected mass distribution are constant with radius. \citet{sonnenfeld13} and \citet{ritondale19a} included external shear in the lens model and used a pixelised source reconstruction. \citet{auger10b} instead described the source using multiple Gaussian and S\'ersic profiles and did not use external shear. It is possible that the lack of external shear leads to biased elongations in case of unexplained deviations from the SIE distribution. We cannot fully determine if these differences create systematic biases: a more consistent choice of models on the observational side would be a benefit for future works. A one-to-one comparison would require creating mock lensing images from simulations and model them as real data, which is beyond the scope of this paper and we leave for future work. 

\section{Results: halo shapes}
\label{sec:Shapes}

In this Section, we present the halo shapes calculated directly from the particle distribution in the four considered scenarios. We discuss how the inclusion of self-interactions and baryons influences the 3D and projected shapes and we attempt a first comparison between simulated and observed values.

\begin{figure*}
    \centering
	\includegraphics[width=\textwidth]{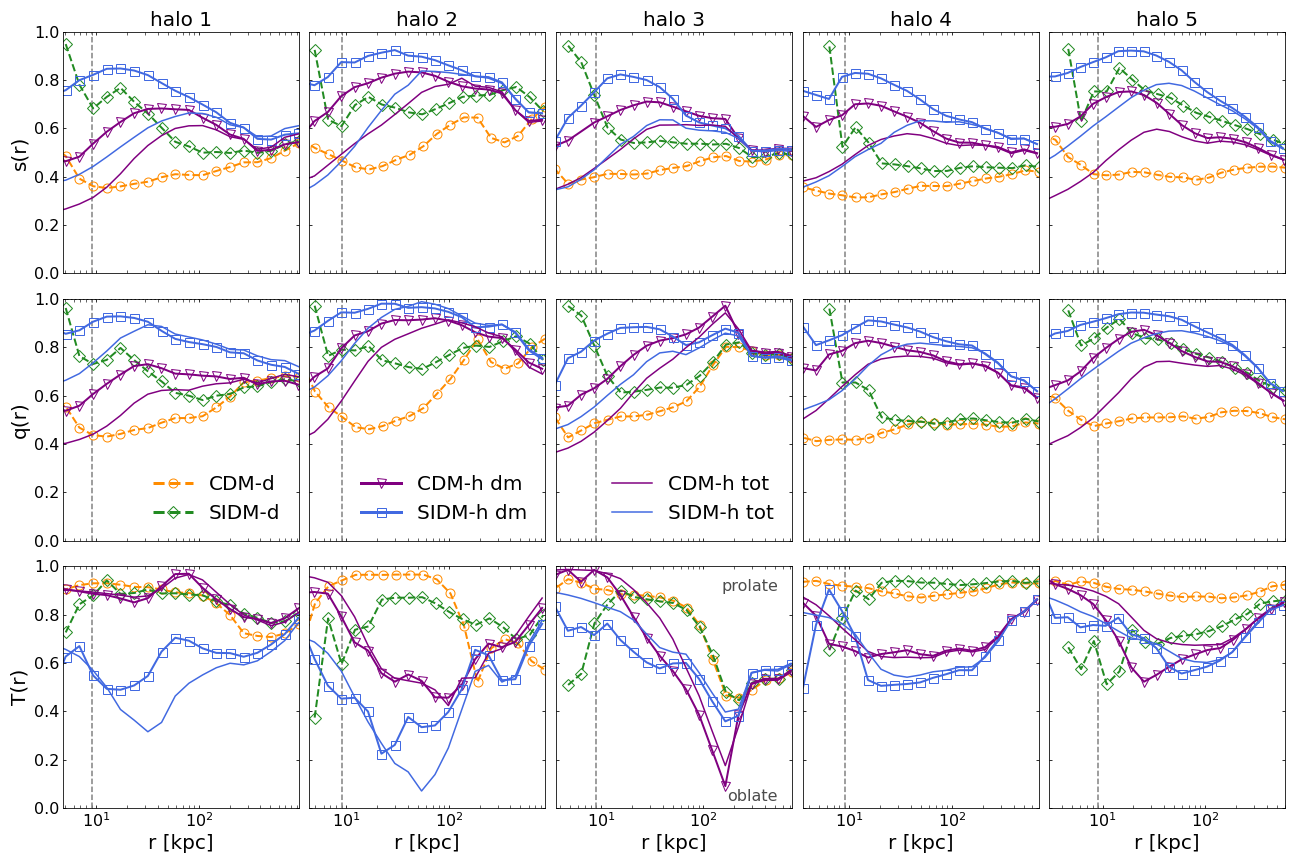}
	\caption{Shapes of the dark matter component and of the total mass distribution as a function of (the equivalent spherical) radius in the range $[0.01,1]\times R_{\rm vir}$ for each halo (columns) and simulation; in the top, middle and bottom panels we show the two axial ratios $s$ and $q$, and the triaxiality $T$. The dark matter shapes in the CDM-d, SIDM-d, CDM-h and SIDM-h simulations are represented respectively by orange, green, blue and purple lines together with circles, diamonds, triangles and squares. For the hydro runs, the two solid lines show instead the shape of the total mass distribution. We require a minimum of 1000 particles in each bin to calculate the shapes; the vertical dashed line shows the convergence radius calculating for shapes following \citet{chua19}, equal to 9$\epsilon$. This radius - well above the spatial resolution - marks the distance from the halo centre from which the shape values can be considered reliable.}
	\label{fig:3Ddm}
\end{figure*}

\begin{figure*}
    \centering

	\includegraphics[width=\textwidth]{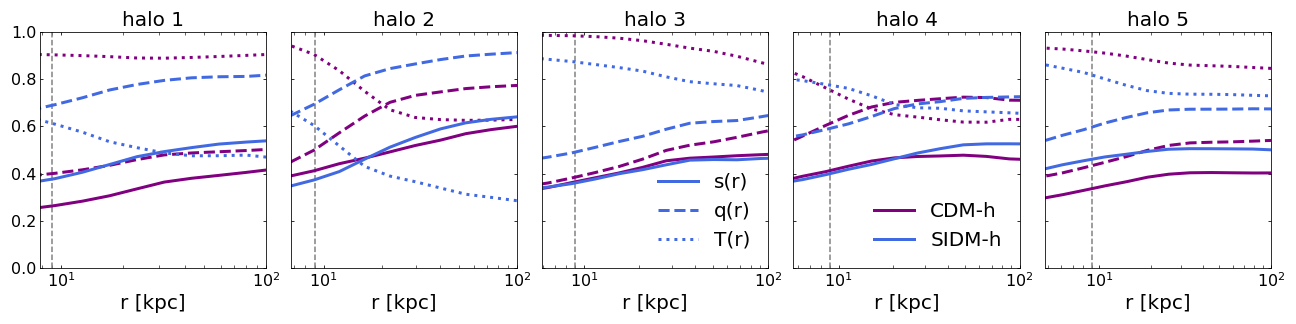}
	\caption{3D shapes of the stellar component in the full physics runs as a function of radius for each halo, in the inner region of the halo between $0.01\times R_{\text{vir}}$ and 100 kpc. Solid (dashed, dotted) lines show the axial ratio $s$ ($q$, $T$) in the CDM-h (purple) and SIDM-h (blue) hydro runs. The vertical dashed line corresponds to the shape convergence radius from Figure \ref{fig:3Ddm}.} 
	\label{fig:3Dbar}
\end{figure*}

\begin{figure*}
	\includegraphics[width=\textwidth]{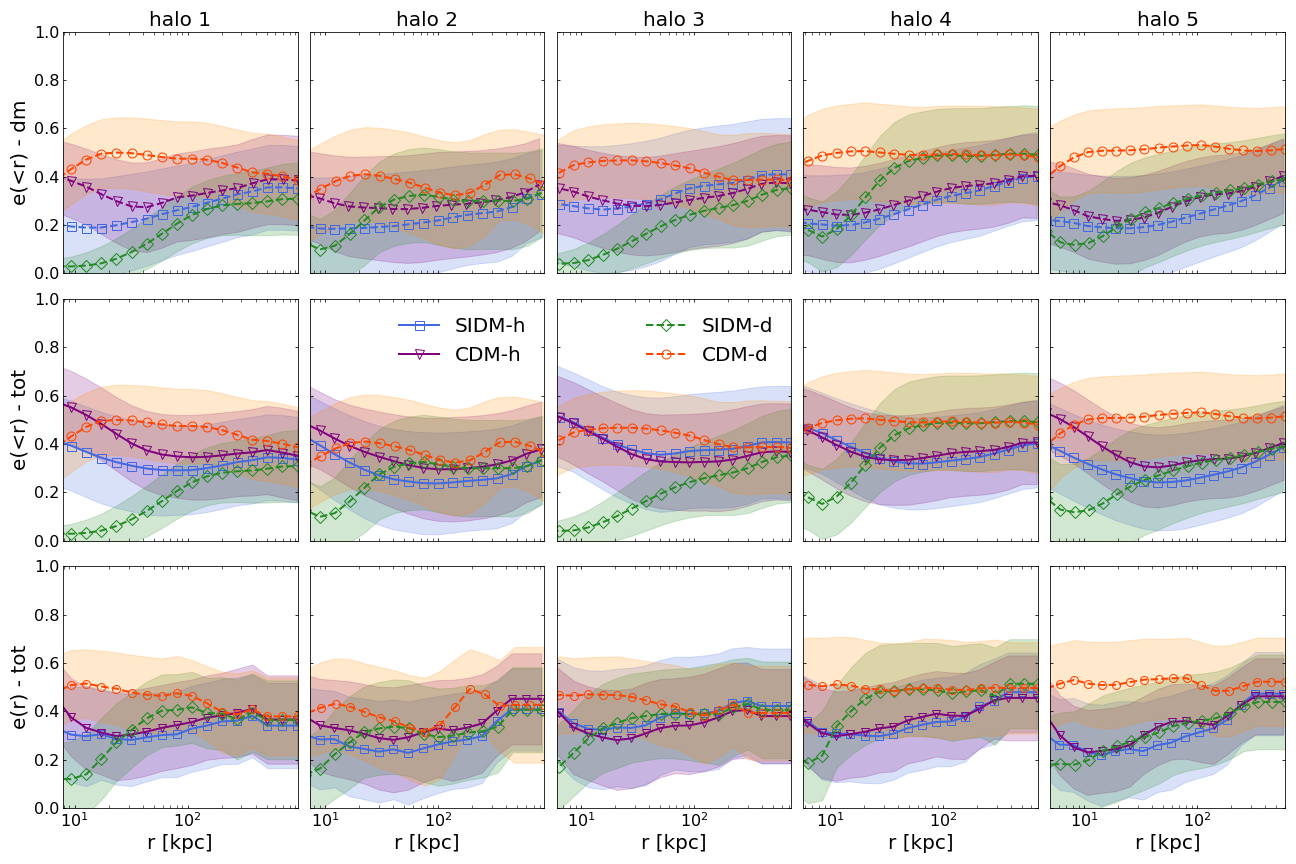}
	\caption{Projected shapes. In all panels we show the mean ellipticity profile and associated 1$\sigma$ region (shaded) for each halo (columns) and model (colours and symbols, same as Figure \ref{fig:3Ddm}). For each halo, we calculated projected shapes in 30  projections and calculate the mean distribution for the projected mass enclosed within an ellipse of radius $r$ in the top and middle panels - and for the projected mass inside shells of the same radius in the bottom panels. The top panels show the shape of the dark matter component, while the middle and bottom one consider the entire mass distribution. The considered radial range is the same of Figure \ref{fig:3Ddm}.}
	\label{fig:2Davg}
\end{figure*}

\subsection{Triaxial shapes in 3D}    
\label{subsec:3dresults}

We start by measuring the intrinsic three-dimensional shape of each halo in our sample. First, we look at the shape of the dark matter component alone in each scenario, to highlight the effect that the presence of baryons has on the underlying distribution of dark matter. The radial profiles of the axis ratios $s$ and $q$ and the triaxiality parameter $T$ are shown in Figure \ref{fig:3Ddm}, for each halo (columns) and simulation (different colour and associated symbol). At each radius the halo shape is defined by a triaxial ellipsoid, calculated iteratively by deforming the initial spherical volume to find the best match to the mass distribution (see Section \ref{sec:Methods} for more details). For simplicity, in Figure \ref{fig:3Ddm} and in general in all figures in this work, we plot the radial shape profiles as a function of the equivalent spherical radius, i.e. the initial sphere with the same volume of the best-fit ellipsoid at that radius.

We recover the well-known trend in CDM dark-matter-only simulations, where the inner parts of haloes are more elongated (i.e. have smaller axis ratios) than the outskirts \citep{allgood06,despali17,chua19}: the inner shape of the halo preserves a trace of the halo merging history and the main direction of accretion, while the outer parts become rounder by subsequent interactions with the field and neighbouring systems. All CDM-d haloes have inner prolate shapes ($T\rightarrow1$ and $s\sim q$) and thus, in the approximation by \citet{Peter:2013II} classify as prolate objects with $a>b=c$ (see Section \ref{sec:Methods}). This picture is modified both in the presence of self-interactions, baryonic physics, or both together. In the absence of baryons (SIDM-d runs), the interactions between the dark matter particles modify the central shape  and the inner parts of the halo become rounder: the same exchange of energy and momentum that creates the central core also partially erases the central triaxiality and its preferential direction \citep{Peter:2013II}. This is evident in Figure \ref{fig:3Ddm} by comparing the SIDM-d and CDM-d case: with self-interactions, the axis ratios increase towards a spherical case in the inner regions, while the shapes tend to coincide at outer radii.

The inclusion of baryons provides another mechanism to generate halo shapes that are rounder than in the dark-matter-only case both in CDM-h and SIDM-h haloes. Moreover, the overall trend of halo shape with radius is reversed and the shapes in the outer parts of the haloes are now more elongated than the centre. This can be easily understood by looking at  the dark matter fraction as a function of radius (right panel in Figure \ref{fig:frac}): the baryons dominate in the innermost $\sim$30 kpc, while at larger radii dark matter determines the halo properties - where at the same time the lower density makes self-interactions less frequent - and the halo shapes converge to similar values in all models.
In CDM-h, our results are compatible with those from previous works \citep{chua19} and in SIDM-h we find an even more pronounced central sphericity, due to the combined effect of baryons and self-interactions \citep[see also ][]{shen22,robertson21}. In both cases, and especially in SIDM-h, the haloes move away from a clear prolate shape ($T=1$) and become more oblate.

We then turn our attention to the baryons and measure the shape of the total mass distribution. The two solid lines (without symbols) show the radial shape of the total mass distribution for the hydro runs (while it is identical in the dark-matter-only runs). In CDM-h and SIDM-h, the total mass distribution is more elongated at the centre than the dark matter halo alone due to the contribution of the galaxy. We also note how, when we look at the total mass distribution, the SIDM-d haloes are clearly the roundest at the centre. Finally, we measure separately the shape of the stellar component - that dominates at the very centre of haloes. All our haloes host early-type galaxies in both dark matter scenarios - see the stellar masses in Table \ref{tab:HaloID} and a more detailed description of the galaxy morphologies in \citet{Despali:2019}. For this reason, we expect these shapes to be quite regular, as the galaxies do not have compact very triaxial components as spiral galaxies. This is indeed the case and the stellar distribution displays a mild, but regular trend with radius in all cases, as can be seen in Figure \ref{fig:3Dbar} where we show the shape profiles of the stellar component.
However, it is evident that, not only the halo shape is rounder in SIDM-h, but also the central elliptical galaxy: both axis ratios are on average higher in the SIDM case, supporting the conclusion that the effects self-interactions and baryons push each other in the direction of rounder central shapes. 

From the inertia tensor, we also derive the orientation of the best-fit ellipsoid (via the eigenvectors) and the misalignment between the dark matter and stellar shapes, in terms of the angle between the major axes. We find misalignments between the two components in agreement with previous works \citep{velliscig15} and that the misalignment is mostly due to the twists of the dark matter distribution as a function of radius - while the stellar distribution maintains an almost constant orientation at all radii (within $\sim$15 degrees in both dark matter scenarios).

\subsection{Projected shapes}

In reality, most astrophysical observations cannot measure the three-dimensional shapes of haloes and galaxies, but only projected quantities on the plane of the sky. Given that haloes are triaxial, the projected shape strongly depends on the viewing angle. Thus, to get a statistically relevant measurement, we rotate each halo 30 times and then project the mass distribution along the three principal axes to calculate the projected 2D shape (see Section \ref{sec:Methods}). This results in a sample of 3$\times$30 measurements for each halo; we then average together different projections in order to derive a distribution of observed ellipticity.

The results are shown in Figure \ref{fig:2Davg}, where we plot the average projected ellipticity $e=1-b/a$ as a function of (the spherical) radius for each system (columns) and simulation (different colors, matching Figure \ref{fig:3Ddm}). We calculate projected shapes for the dark matter component alone (first now) and for the total mass distribution (second and third row) - these are identical for the CDM-d and SIDM-d runs and differ only in the runs containing baryonic physics. Moreover, we calculate the shapes of the total mass distribution both by considering the enclosed mass at each radius (second row) and the mass in ellipsoidal shells of the same radius (bottom row). 
For each halo and considered case, we show the mean value of the measured ellipticity together with the 1$\sigma$ region (shaded area). 
By comparing the first row of Figure \ref{fig:2Davg} to Figure \ref{fig:3Ddm}, we see that - on average - projected dark matter shapes are less elongated than triaxial ones and that the projection reduces the differences between different models \citep{despali17}. The outer parts of the haloes (similar in all models) lie close to the centre in projection and contribute to the shapes. Projection effects also greatly reduce the misalignment between the dark matter and stellar component: we calculate the orientation of the main axes of the best-fit ellipse, by means of the eigenvectors corresponding to the longest axes of both component at each radius, and find an average misalignment of $5\pm5$ degrees in CDM-h and $7\pm10$ degrees in SIDM-h. For the latter, rounder shapes probably introduce a larger degree of uncertainty in the orientation of the longest axis.

When we consider the total mass distribution (second and third row in Figure \ref{fig:2Davg}), we find that the projected ellipticity increases at the centre due to the galaxy shape, consistently with Figure \ref{fig:3Ddm} and \ref{fig:3Dbar}, setting the SIDM-d distribution apart from the other three cases, especially when considering enclosed shapes $e(<r)$ as in the top and middle row of Figure \ref{fig:2Davg}. 

This demonstrates once again how the inclusion of baryons is essential to derive realistic predictions of the observed properties of galaxies and haloes in the real universe, while using dark-matter-only simulations likely leads to overestimate the differences between CDM and SIDM. A similar effect is found by \citet{robertson21} and Mastromarino et al. \emph{in prep} when analysing the halo density profiles of a large number of haloes in SIDM and CDM.

\subsection{Comparison with observational measurements} \label{observations}

\begin{figure}
\centering

    \includegraphics[width=\hsize]{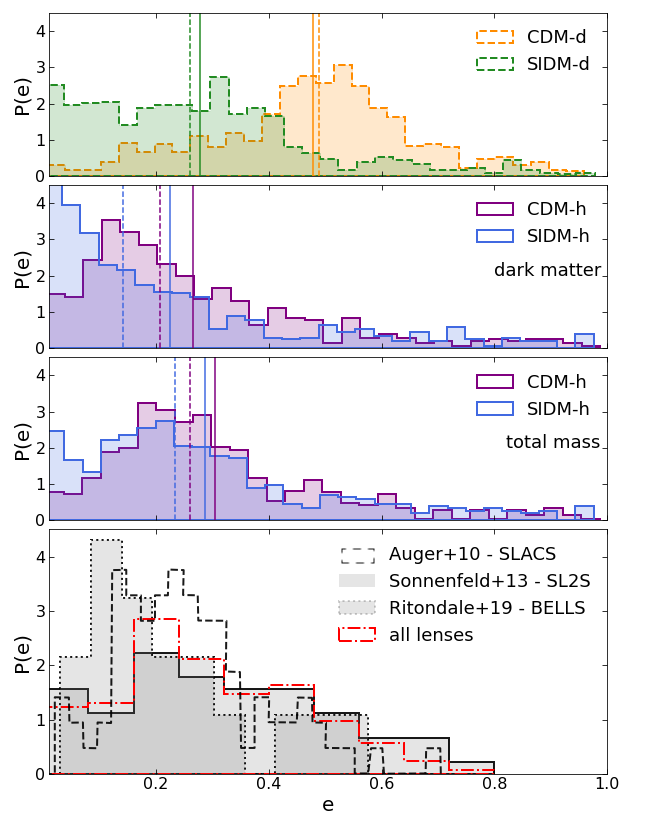}
	\caption{Normalised distribution of projected ellipticity at the centre of the haloes. For simulated values (first three rows), we plot the ellipticity at a distance from the halo centre within r = 14 kpc. The top panel shows the distributions derived from the dark-matter-only runs, whereas for the hydro runs, we show the results calculate both by using the dark matter particles only (second row) and dark matter plus baryons together (third row). In each panel, the vertical solid (dashed) lines of the corresponding color show the mean  (median) of each distribution. Finally in the bottom panel, we report the observational distributions derived from gravitational lenses in the SLACS \citep{auger10b}, BELLS \citep{ritondale19a} and SL2S \citep{sonnenfeld13} samples. The red histogram shows the normalised distribution of the total sample of available lenses.
	}
	\label{fig:2Dhist}
\end{figure}

\begin{figure}
\centering
    \includegraphics[width=0.98\hsize]{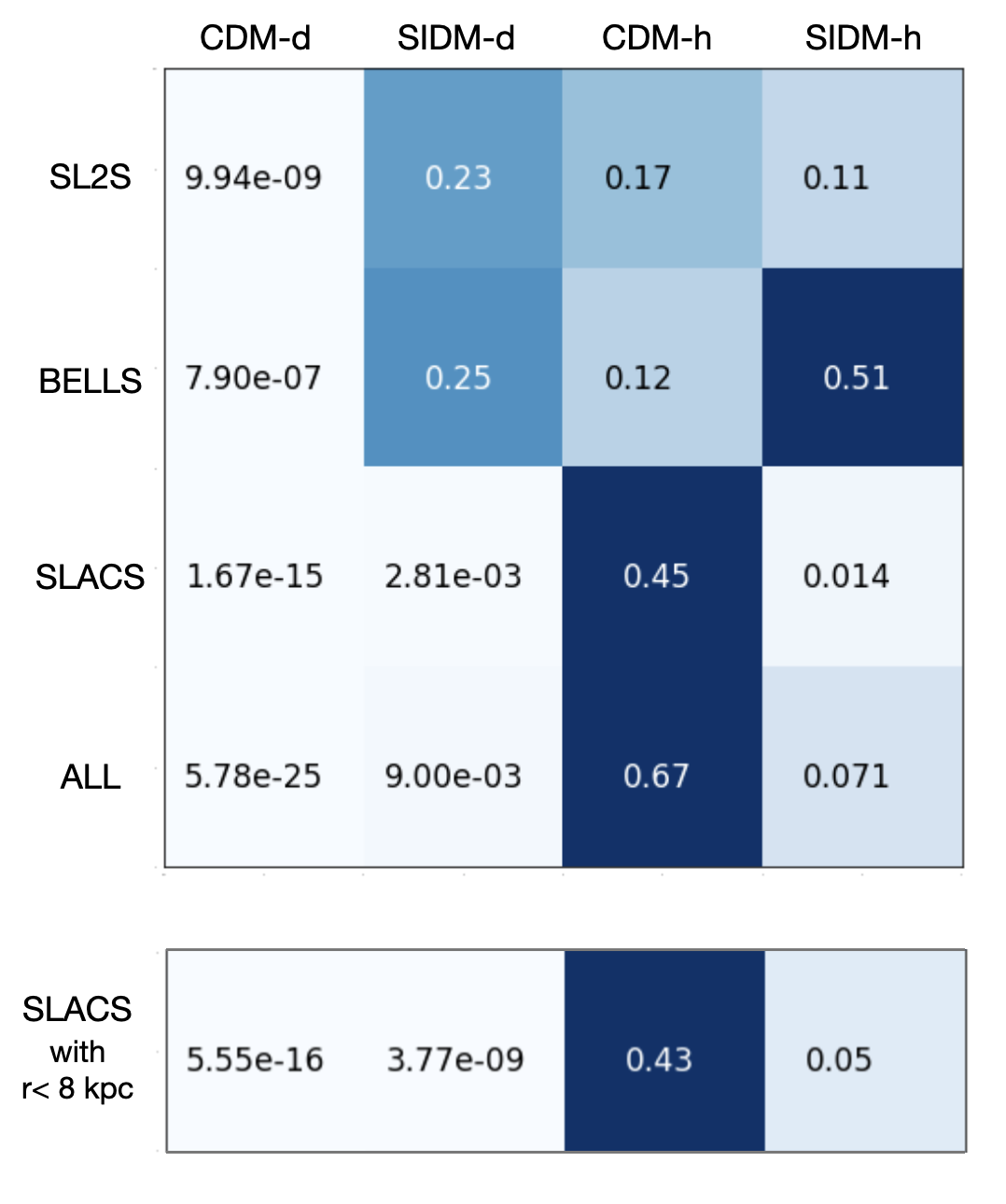}
	\caption{Results of the KS test for each pair of data-sets. 
	We compare the simulated samples within $r=14$ kpc (columns) with observations (rows). For each pair show the $p$-value: when the $p$-value is small (light blue), two samples are unlikely to be drawn from the same parent distribution. In the bottom panel we compare the SLACS lenses with the simulated shapes measured at a smaller radius $r=8$ kpc, closer to the average size of the lensed images.}
	\label{fig:pval}
\end{figure}

\begin{figure}
\centering

    \includegraphics[width=\hsize]{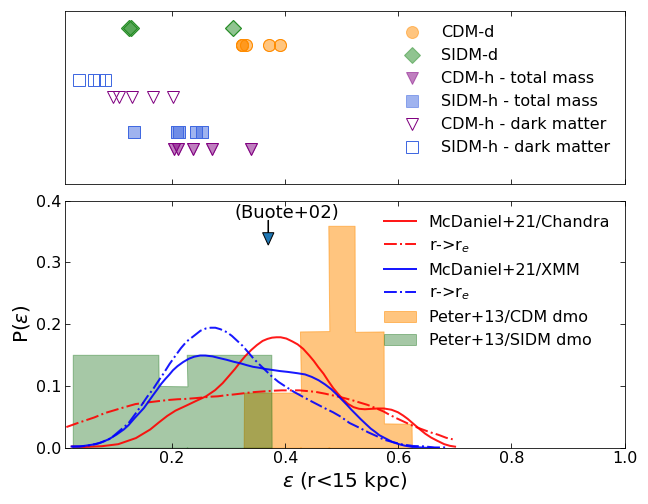}
	\caption{Distribution of \emph{spheroidal} ellipticity parameter $\epsilon$ at a distance from the halo centre r < 15 kpc. Since this is a measure of the 3D shape, we only have five values (one per halo) per model, shown in the top panel. These are calculated from the triaxial axes as described in Section \ref{sec:Methods} and following \citet{Peter:2013II}. In the bottom panel, we report the observational distributions from \citet{mcdaniel21} for \emph{Chandra} and \emph{XMM-Newton} data and the result from \citet{buote02} on NGC 720, as well as the distributions from the dark-matter-only simulations from \citet{Peter:2013II}. CDM and SIDM with $\sigma_T/m_{\chi}=1 $cm$^{2}$/g are shown in orange and green, respectively.}
	\label{fig:2Dhist2}
\end{figure}

We now attempt a comparison between the projected ellipticity measured from our simulations and observational values coming from: $(i)$ the modelling of gravitational lenses from the SLACS \citep[][80 lenses]{auger10b}, BELLS \citep[][16 lenses]{ritondale19a} and SL2S \citep[][56 lenses]{sonnenfeld13} samples and $(ii)$ the analysis of X-ray emission of elliptical galaxies from \citet[][one galaxy]{buote02} and \citet[][11 galaxies]{,mcdaniel21}, where the projected shape of the potential is recovered by matching the X-ray surface brightness distribution to predictions (see Section \ref{sec:Methods} for more details).
In both cases, we restrict the measurement of projected shapes to the inner part of the haloes in order to compare with observations, choosing the appropriate radii.

In Figure \ref{fig:2Dhist} we show the normalised distribution of projected ellipticities $e$ calculated within r = 14 kpc, which we choose as an estimate of the area that is probed by lensing. This choice is somewhat arbitrary, given that the mass profile used for the lens modelling has a constant ellipticity, and comparisons with a different radius could lead to slightly different results. We discuss this aspect further in Appendix \ref{sec:App1}. For comparison, the mean Einstein radii (i.e. approximately the size of the lensed images) are 5.3 kpc, 12.8 and 13.2 kpc, respectively for the SLACS, BELLS, and SL2S samples. The SLACS lenses are thus on average smaller, which might influence our results. In Figure \ref{fig:2Dhist}, we compare all observations to the simulated values measured within 14 kpc, but later in this Section we also use shapes within a smaller radius for the SLACS sample, finding consistent results.

The dashed histograms in the top panel show the results for the  dark-matter-only runs, while the second and third panel show the ellipticity distribution in the simulations including baryons.
The ellipticity distributions differ significantly in the dark-matter-only runs, reproducing previous results \citep{Peter:2013II}. However, when it comes to the runs including baryons the  two distributions are closer, especially when the total mass distribution is considered (third row) instead of the dark matter component alone (second row). We note, however, that the SIDM-h distribution presents a higher tail at very low ellipticities $e\simeq0$, which is not present in the CDM-h case: its presence (or absence) could be used to distinguish between the two models, instead of focusing only on the values.

Finally, the bottom panel summarises the normalised ellipticity distribution from observational lensing works: the solid, dashed and dotted black histograms report the projected ellipticity from the three considered samples. We remind the reader that each sample has a different size (see Section \ref{sec:Methods3}): the SLACS sample is the largest (80 lenses), followed by SL2S (56) and BELLS (16). In order to reduce potential selection biases, we also calculate the normalised distribution of the total sample of lenses, shown here by the red dot-dashed histogram. We can already see by eye that none of them is compatible with the distribution predicted by the CDM-d dark matter-only case (orange histogram), whereas they look close to the hydro results. As mentioned in Section \ref{sec:Methods}, we have not gone through the process of creating mock images from the simulations and modelling them as real data, which we leave for future work, and thus the comparison could be biased. However, we impose an elliptical shape that can adapt to the mass distribution in terms of orientation and ellipticity (similar to the elliptical profile used in observations) and we use only the particles belonging to the main halo - this should reduce the need of accounting for a possible source of shear in our simulated data. We refer to Appendix \ref{sec:App1} for more details. 

We use the Kolmogorov-Smirnov (KS) test to quantitatively compare the simulated and observed lensing samples: we test their compatibility against the null hypothesis that they are sampled from the same parent distribution. The tables in Figure \ref{fig:pval} report the results of the test for each combination of samples in terms of the $p$-value. This is the probability of observing this maximum difference (KS distance) found between the cumulative distribution functions of the two samples, under the assumption that the parent distribution is the same. If the KS statistic is small or the $p$-value is high, then we cannot reject the null hypothesis in favour of the alternative. All observational samples clearly disfavour the high ellipticities from the CDM-d run, with extremely low $p$-values  (first column in Figure \ref{fig:pval}). What is evident is that the data reject the CDM-d predictions and require the presence of a component that can increase the central sphericity in order to explain the distribution of observed shapes.

The SLACS sample favours the CDM-h case, clearly rejecting the two dark runs, while the BELLS lenses show a certain degree of preference for the SIDM-h run. The SL2S lenses show a slightly higher preference for the SIDM-d distribution, but not much higher than CDM-h. The other combinations are less likely, but cannot be fully ruled out, and in general, the degree of preference for each sample could be due to the different techniques used to obtain observed and simulated values. When considering the total sample of 152 lenses together (last row), the gravitational lenses prefer the CDM-h scenario, followed by SIDM-h, and rule out the dark runs. This result is intriguing but has to be interpreted with caution for a number of reasons: $(i)$ the clear preference of the (most numerous) SLACS sample for CDM-h dominates the constraints; $(ii)$ at the same time, the lens modelling of the BELLS and SL2S samples includes a higher level of complexity (see Section \ref{sec:Methods3}) and  both hydro runs can partially explain the BELLS and SL2S, even if to a different degree; $(iii)$ we have calculated the projected shapes along many independent projections, but the halo sample is limited; $(iv)$ differences in the details of the lens modelling of the three samples could influence the results. We thus argue that larger (and more complete) samples of observed and simulated galaxies will be needed to confirm this result.

The comparison done here is based on the shapes measured within $r=14$ kpc: in Appendix \ref{sec:App1}, we show the KS and $p$-values trends with radius, and we demonstrate that our conclusions are valid even for a different choice of radius up to 40 kpc from the centre, while discussing other possible sources of bias. Finally, the lower panel in Figure \ref{fig:pval} shows the $p$-values of the comparison between the SLACS sample with the simulated shapes at r < 8 kpc, a smaller radius closer to the size of the SLACS Einstein radii, that leads to similar result.

We also compared the observed samples and the simulated ones separately among each other: we found that the KS test is able to distinguish between the different simulated samples and that, conversely, observed lenses are mostly compatible with each other. For sake of clarity, we expand the discussion in Appendix \ref{sec:App1}  and show the $p$-values of these additional comparison in Figure \ref{fig:pval2} and \ref{fig:app1}.

Finally, we compare the projected shapes with results from X-ray studies. In the next Section, we expand this comparison by analysing mock X-ray images. We remind the reader that the definition of ellipticities shown so far differ slightly from those reported by \citet{buote02} and \citet{Peter:2013II}, because we plot the projected ellipticities $e$ and not the \emph{spheroidal} ellipticities $\epsilon$ (see Section \ref{sec:Methods});  given our small sample of haloes, we cannot obtain a full distribution of the latter, but only one value per halo. However, we calculate the $\epsilon$ value (see Section \ref{sec:Methods}) of each halo at r < 15 kpc (consistently with the observational analysis from previous works) and show a comparison in Figure \ref{fig:2Dhist2}. In practice, all our haloes are prolate in the inner parts and thus we use the three axes of the best-fitting ellipsoid to calculate $\epsilon = 1-\sqrt{bc}/a$, corresponding to the points in the top panel. In the bottom panel, the dashed and solid curves report the observational results from \citet{mcdaniel21}, who calculated the distribution of \emph{spheroidal} ellipticity of haloes that would better reproduce the observed ellipticity profiles. For comparison, the coloured histograms show the results from dark-matter-only simulations from \citet{Peter:2013II}, who used the same method to calculate $\epsilon$ from the 3D shapes.
Unfortunately in this case, we cannot draw significant conclusions, as our statistics are limited; we will discuss a more meaningful comparison with this data set in the next Section, where we analyse the shape profiles and not limit the comparison to a specific radius.

\section{Results: shapes of X-ray images}   \label{sec:Xray}

\begin{figure*}
	\includegraphics[width=0.85\hsize]{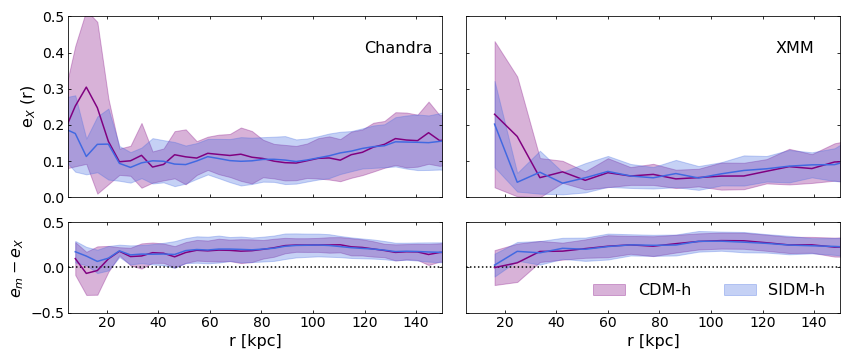}
     \caption{Shapes of the X-ray emission, measured in ellipsoidal shells by calculating the inertia tensor of the mock images, created to match \emph{Chandra} (left) and \emph{XMM-Newton} (right) observations. 
     The solid lines and the shaded area stand for the mean and standard deviation of the ellipticity profile $e_{X}$ as a function of radius. In the bottom panel, we show the mean difference between the ellipticity calculated from the projected mass density in Section \ref{sec:Shapes} and that of the X-ray emission. The first describes the mass distribution, while the second traces the gravitational potential and is rounder by construction.}
    \label{fig:ShapeXray}
\end{figure*}

\begin{figure}
	\includegraphics[width=0.95\hsize]{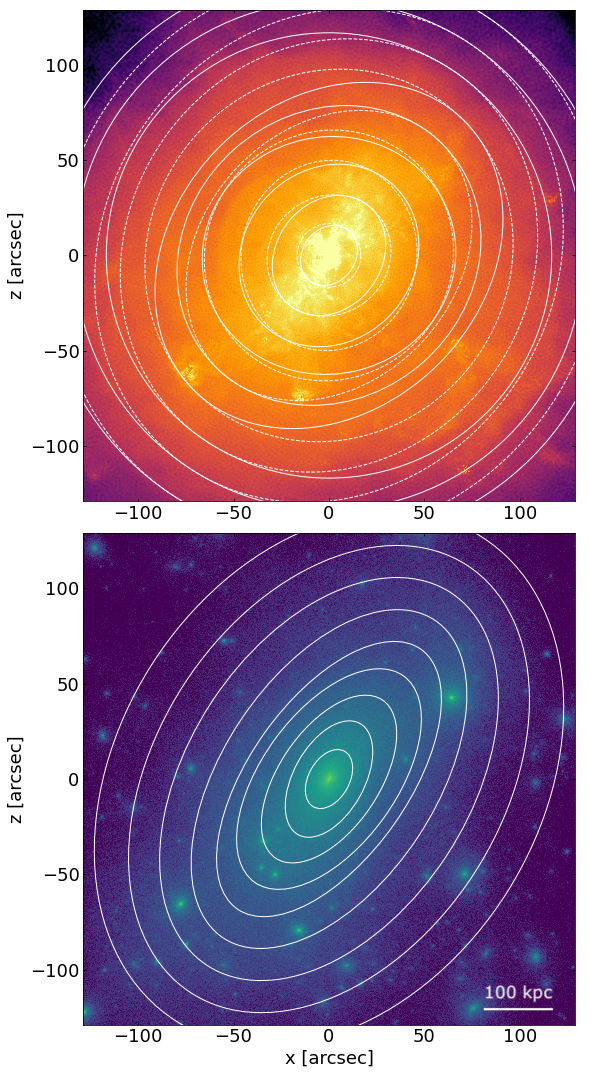}
     \caption{Halo shapes from the X-ray emission vs. projected mass distribution. In the top panel we show the \emph{Chandra} simulated emission for a projection of halo 1 (see Figure \ref{fig:MockXray}, together with the best-fit elliptical shells estimated by measuring the inertia tensor of the image (solid white line). The dashed lines show instead the ellipticity estimated from the corresponding \emph{XMM-Newton} mock image, at the same radii. In the bottom panel, we show instead the projected dark matter density, together with the elliptical shells that best fit the projected mass distribution. These more elongated than the ones in the top panel, as predicted by the difference between the shape of the mass and the potential. They have, however, similar orientation. }
    \label{fig:ShapeXray3}
\end{figure}

\begin{figure}
	\includegraphics[width=\hsize]{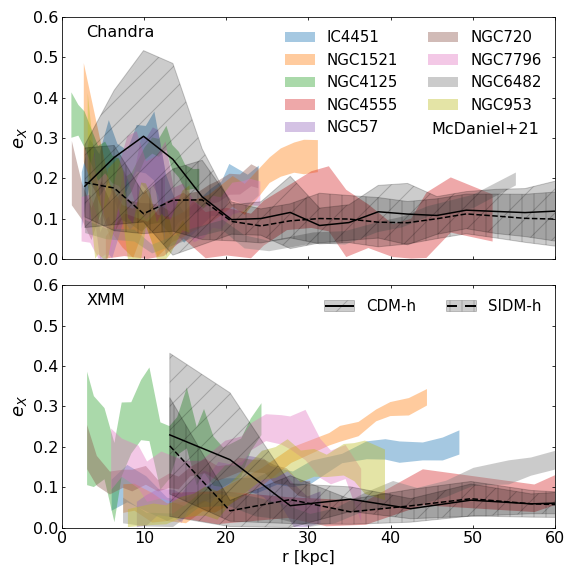}
     \caption{Comparison between the mean ellipticity profiles derived from the CDM-h (solid line) and SIDM-h (dashed line) runs and the measurements for nine elliptical galaxies from from \citet{mcdaniel21}, shown by bands of different colours. }
    \label{fig:ShapeXray2}
\end{figure}

We use the pipeline from \citep{Barnes2020} to create mock X-ray luminosity maps, as described in Section \ref{sec:Methods} and shown in Figure \ref{fig:MockXray} and \ref{fig:MockXray2}. We generate 90 maps for each halo by rotating the particle distribution, similarly to the case of projected 2D shapes. In this case, we use the total gas distribution in the halo (and not only that belonging to the main subhalo).

We measure the ellipticity of the X-ray emission in each map by calculating the inertia tensor directly on the image, following \citet{buote02,mcdaniel21}, as described in Section \ref{sec:Methods}.
We then calculate the mean ellipticity for the CDM-h and SIDM-h cases. Figure \ref{fig:ShapeXray} shows the X-ray ellipticity $e_{X}(r)$ in the \emph{Chandra} (left) and \emph{XMM Newton} (right) mock images. Thin purple (blue) lines show the results for each halo in the CDM-h (SIDM-h) scenarios, whereas the thick lines of the corresponding colour stand for the mean, together with the associated standard deviation. 
At the very centre, the CDM-h shapes are more elongated (but still compatible within the shaded regions), but this effect quickly disappears at distances $r\geq$ 25 kpc and the ellipticity profile becomes flat. 
Our results are consistent with the cluster-scale measurements from \citet{shen22}, who also found a maximum mean difference of $\simeq$0.1 in the central ellipticity (in their case, within 0.1$\times R_{\text{vir}}$) which then disappears further away from the halo centre. \emph{XMM-Newton} shapes are rounder than the corresponding \emph{Chandra} measurements due to the lower resolution that smooths out part of the elongation.

In the bottom panels, we compare - at each radius - the ellipticities $e_{X}$ with those obtained in Section \ref{sec:Shapes} by modelling the projected mass distribution (see Figure \ref{fig:2Davg}). The mean ellipticities inferred from the mock X-ray images are lower than those measured from the projected mass distribution. This is consistent with the fact that the gas - under the hypothesis of hydrostatic equilibrium - traces the gravitational potential, which is rounder than the mass distribution by construction \citep{golse02}. We show one visual example of these differences in Figure \ref{fig:ShapeXray3}. 

Apart from this difference, the X-ray ellipticities in CDM-h and SIDM-h are even closer to each other than the values calculated from the projected mass difference: it seems unlikely that we will be able to distinguish between the two dark matter scenarios only by measuring the shape of the X-ray emitting gas. However, it is encouraging that the simulated images are a good representation of real observations: in Figure \ref{fig:ShapeXray2} we compare the mean ellipticity profiles to the observational profiles measured by \citet{mcdaniel21} for nine galaxies, using \emph{Chandra} and \emph{XMM-Newton} data in the same band. We only use the systems for which they report observations with both instruments. Despite the object-to-object variation and the difference in redshift between simulations and observations, the overall trend is well reproduced by the \emph{Chandra} mocks. In the case of \emph{XMM-Newton}, some observed systems display an increase of ellipticity at $r>30$ kpc that we do not see so strongly in our simulations. However, given the small sizes of both samples, we cannot judge whether this is a real discrepancy or an effect of the limited statistics. 

Based on this comparison, we cannot discard one of the two dark matter scenarios, nor establish if one of the two should be preferred: both the CDM model and a SIDM model with constant cross-section of $\sigma_T/m_{\chi}=1 $cm$^{2}$/g are consistent with the data.  This means that previous constraints on the self-interaction cross-section $\sigma$ might overestimate our ability to distinguish between these two models only by looking at the shapes.

Finally, we note that an important difference between the observational analysis and the simulated mock images is that we have assumed that the latter are noise-free and that the background is irrelevant, while this is obviously not the case for observational data.

\section{SUMMARY and DISCUSSION}
\label{sec:summary}

Using zoom-in re-simulations of a sample of massive galaxies at $z=0.2$ \citep[see][]{Despali:2019}, we investigate the possibility of discriminating between CDM and a self-interacting dark matter model with constant cross-section $\sigma/m =1 $cm$^{2}$/g on the basis of halo shapes. To this end, we measure the three-dimensional and projected shapes of the dark matter and total mass distribution of five systems, simulated in the two considered scenarios, with (CDM-h and SIDM-h) or without (CDM-d and SIDM-d) the inclusion of baryonic physics - thus a total of four runs per halo. Our main results are:

\begin{itemize}
    \item we recover the well-known trend in CDM dark-matter-only simulations, where the inner parts of the haloes are more elongated than the outskirts;
    \item similarly, we reproduce the results of previous works on SIDM dark-matter-only simulations, where the self-interactions in the high-density centre of the halo produce rounder shapes;
    \item we find that the inclusion of baryons on average leads to rounder shapes both in the CDM and SIDM scenarios and that the predictions from the two dark matter models are very similar when hydrodynamical simulations are used, given that baryons dominate in the inner region;
    \item the effect of both baryons and self-interactions influences the shape of the inner part of the halo, while for each halo the shapes converge to the same values at large radii in all scenarios: far from the centre the dark matter dominates the density, but at the same time the density is low and thus self-interactions are less frequent than in the centre;
    \item we measure projected shapes along 90 random viewing angles for each system and calculate the ellipticity of the projected mass distribution: the mean projected shapes are closer to spherical and more difficult to distinguish compared to the three-dimensional measurements, due to the contribution of the outer part of the haloes that fall into the considered projected radius.
\end{itemize}

We then compare the distribution of simulated shapes to observational results in the context of gravitational lensing \citep{auger10b,sonnenfeld13,ritondale19a} and X-ray \citep{buote02,mcdaniel21} studies. In the case of gravitational lensing, we compare the projected shapes - measured in the simulations from the mass density distribution within 14 kpc - to the ellipticity of the best SIE profile that was measured from the lensing convergence. We use three lens samples: the SLACS \citep{bolton06,auger10b}, the BELLS \citep{ritondale19a} and the SL2S \citep{sonnenfeld13} lenses. 

We perform a Kolmogorov-Smirnoff test to determine whether or not the observed samples are drawn from the same distribution of one of the simulations. We find that:
\begin{itemize}
    \item all the lens samples strongly reject the CDM-d model, which produces too high elongations; 
    \item the SLACS (BELLS) sample prefers the simulated results from the CDM-h (SIDM-h) run. However, none of the individual samples clearly rejects the other simulated results (from CDM-h, SIDM-h and SIDM-d);
    \item if we consider all observed lenses together, the observational data favour the CDM-h case,;
    \item the trends are valid for most radii in the range between 5 and 30 kpc and the CDM-h (followed by SIDM-h) scenario provides the best overall match at all radii (see Appendix \ref{sec:App1});
    \item when applied to the simulated values only, the KS test is able to distinguish between the samples. Conversely, the observed samples are compatible to each other (even though at different degrees, see Appendix \ref{sec:App1}).
\end{itemize}

The preference of the lensing data for the CDM-h scenario is intriguing, but has to be taken with caution due to the small number of simulated haloes, the potential differences between the measurements in simulations and observations, and the difference in the modelling of the observational samples (see Section \ref{sec:Methods3} and \ref{observations}) A larger set of simulated haloes and a more consistent modelling technique of all the lensing data would be needed to confirm to confirm this result. 

Finally, we create mock X-ray maps from the simulations including baryonic physics, using the method described in \citet{Barnes2020}. We simulate \emph{Chandra} and \emph{XMM-Newton} observations in the soft X-ray energies between 0.5 and 2 keV, by rotating each halo along different viewing angle, as done with the projected mass distribution. We measure the shapes directly from the imaging, by applying the same procedure used to evaluate the inertia tensor on the image pixels and fluxes \citep{buote02,mcdaniel21}. Under the assumption of hydrostatic equilibrium, the X-ray emission describes the gas distribution and thus directly traces the gravitational potential of the halo. We analysed the ellipticity profiles and compared them with those of observed elliptical galaxies from \citet{mcdaniel21}, in the same band and at similar resolution.
We find that:
\begin{itemize}
\item the ellipticity profiles $e_{X}$ are flat at large radii (r>25 kpc), with an average ellipticity of $\simeq$0.15; 
\item towards the centre, the X-ray ellipticity increases and CDM-h shapes can be larger than SIDM-h ones by a factor of two, but still compatible within 1$\sigma$;
\item the X-ray ellipticities are smaller ($\sim$half) than those calculated from the projected mass distribution at the same radius, as expected by the fact that the gravitational potential is rounder than the mass distribution;
\item the ellipticity profiles are compatible with the \emph{Chandra} and \emph{XMM-Newton} observational results from \citet{mcdaniel21} both for the CDM-h and SIDM-h scenarios.
\end{itemize}

In summary, once baryonic physics is properly accounted for, the shapes of CDM and SIDM systems are much more similar to each than in dark-matter-only simulations. When compared to observational data, the gravitational lenses show a certain degree of preference for the CDM hydro scenario (but the result depends on the considered sample), while the X-ray analysis cannot distinguish between the two models.

We conclude that both CDM and a SIDM model with a constant cross-section $\sigma/m_\chi =1 $cm$^{2}$/g are still compatible with the most recent observations of elliptical galaxies. 
This is consistent with the conclusions by \citet{mcdaniel21}, whereas other previous works based on the comparison to dark-matter-only simulations \citep{buote02,Peter:2013II} found a preference for SIDM models with $\sigma_{T}/m_{\chi} \simeq0.1$cm$^{2}$/g and disfavoured higher cross-section like the one used in this work. It is thus evident how the inclusion of baryons in the simulated results is of fundamental importance to derive realistic predictions. We compared observational results to a sample of five simulated ETGs: larger samples of simulations will help us to derive more precise predictions, avoid selection biases and base the results on complete samples. Moreover, a more consistent analysis of the lensing data would eliminate potential sources of bias and strengthen the comparison with future simulations. Finally, methods that can decompose the mass distribution into dark matter and baryonic components separately might give a better chance of distinguish CDM and SIDM models.

Here, we have used "classic" SIDM simulations with elastic scattering and constant cross-section, but many other SIDM models are still viable alternatives to CDM, such as models with velocity-dependent cross section \citep{zavala19,robertson21}, inelastic scattering \citep{vogel19} or more complex models including dark radiation \citep{vogel16}.
It is also worth mentioning that the CDM-h and SIDM-h hydro runs have been created without re-calibrating the baryonic physics model - both contain the standard TNG model \citep{pillepich18}. Having run only a sample of zoom-in simulations, we do not know if all the standard galaxy scaling relations would be reproduced in a SIDM cosmological box and if using the same hydro model forces a similarity between the SIDM-h and CDM-h halo shapes. However, this would require a much larger computational effort and our results represent a first step forward towards more realistic predictions.

Our results can be interpreted both in a pessimistic and optimistic light. On one hand, it is less straightforward than previously estimated to discriminate between the two considered models by measuring at the shapes of galaxies and their haloes. On the other hand, both models can well reproduce observed distribution and thus are both viable explanations for the properties of elliptical galaxies, leaving us with alternatives in case one of the two models is challenged by other observations or at a different mass scale.

\section*{ACKNOWLEDGMENTS}
We are grateful to the IllustrisTNG collaboration for allowing the use of the IllustrisTNG model for \citet{Despali:2019}, on which this work is based. We thank David Barnes for sharing his X-ray pipeline with us. GD thanks Luca Di Mascolo for useful discussion about X-ray data and  Ralf Klesssen for his thorough comments on the paper draft. We thank the anonymous referee for the positive feedback on the paper and the useful comments. MS acknowledges support by the European Research Council under ERC-CoG grant CRAGSMAN-646955. JZ acknowledges support by a Grant of Excellence from the Icelandic Research fund (grant number 206930). SV acknowledges funding from the European Research Council (ERC) under the European Union’s Horizon 2020 research and innovation programme (LEDA: grant agreement No 758853). 

\section*{Data availability}
The simulations as well as the codes to analyze them will be made available upon request to the corresponding author. This research made use of the public python packages Astropy \citep{astropy13}, matplotlib \citep{matplotlib}, NumPy \citep{numpy} and SciPy \citep{scipy}.


\bibliographystyle{mnras}
\bibliography{main.bbl} 

\begin{thebibliography}{}
\makeatletter
\relax
\def\mn@urlcharsother{\let\do\@makeother \do\$\do\&\do\#\do\^\do\_\do\%\do\~}
\def\mn@doi{\begingroup\mn@urlcharsother \@ifnextchar [ {\mn@doi@}
  {\mn@doi@[]}}
\def\mn@doi@[#1]#2{\def\@tempa{#1}\ifx\@tempa\@empty \href
  {http://dx.doi.org/#2} {doi:#2}\else \href {http://dx.doi.org/#2} {#1}\fi
  \endgroup}
\def\mn@eprint#1#2{\mn@eprint@#1:#2::\@nil}
\def\mn@eprint@arXiv#1{\href {http://arxiv.org/abs/#1} {{\tt arXiv:#1}}}
\def\mn@eprint@dblp#1{\href {http://dblp.uni-trier.de/rec/bibtex/#1.xml}
  {dblp:#1}}
\def\mn@eprint@#1:#2:#3:#4\@nil{\def\@tempa {#1}\def\@tempb {#2}\def\@tempc
  {#3}\ifx \@tempc \@empty \let \@tempc \@tempb \let \@tempb \@tempa \fi \ifx
  \@tempb \@empty \def\@tempb {arXiv}\fi \@ifundefined
  {mn@eprint@\@tempb}{\@tempb:\@tempc}{\expandafter \expandafter \csname
  mn@eprint@\@tempb\endcsname \expandafter{\@tempc}}}

\bibitem[\protect\citeauthoryear{{Allgood}, {Flores}, {Primack}, {Kravtsov},
  {Wechsler}, {Faltenbacher}  \& {Bullock}}{{Allgood} et~al.}{2006}]{allgood06}
{Allgood} B.,  {Flores} R.~A.,  {Primack} J.~R.,  {Kravtsov} A.~V.,  {Wechsler}
  R.~H.,  {Faltenbacher} A.,   {Bullock} J.~S.,  2006, \mn@doi [\mnras]
  {10.1111/j.1365-2966.2006.10094.x}, \href
  {http://adsabs.harvard.edu/abs/2006MNRAS.367.1781A} {367, 1781}

\bibitem[\protect\citeauthoryear{{Andrade}, {Fuson}, {Gad-Nasr}, {Kong},
  {Minor}, {Roberts}  \& {Kaplinghat}}{{Andrade} et~al.}{2022}]{andrade22}
{Andrade} K.~E.,  {Fuson} J.,  {Gad-Nasr} S.,  {Kong} D.,  {Minor} Q.,
  {Roberts} M.~G.,   {Kaplinghat} M.,  2022, \mn@doi [\mnras]
  {10.1093/mnras/stab3241}, \href
  {https://ui.adsabs.harvard.edu/abs/2022MNRAS.510...54A} {510, 54}

\bibitem[\protect\citeauthoryear{{Astropy Collaboration} et~al.,}{{Astropy
  Collaboration} et~al.}{2013}]{astropy13}
{Astropy Collaboration} et~al., 2013, \mn@doi [\aap]
  {10.1051/0004-6361/201322068}, \href
  {http://adsabs.harvard.edu/abs/2013A%26A...558A..33A} {558, A33}

\bibitem[\protect\citeauthoryear{{Auger}, {Treu}, {Bolton}, {Gavazzi},
  {Koopmans}, {Marshall}, {Moustakas}  \& {Burles}}{{Auger}
  et~al.}{2010}]{auger10b}
{Auger} M.~W.,  {Treu} T.,  {Bolton} A.~S.,  {Gavazzi} R.,  {Koopmans}
  L.~V.~E.,  {Marshall} P.~J.,  {Moustakas} L.~A.,   {Burles} S.,  2010,
  \mn@doi [\apj] {10.1088/0004-637X/724/1/511}, \href
  {https://ui.adsabs.harvard.edu/#abs/2010ApJ...724..511A} {724, 511}

\bibitem[\protect\citeauthoryear{Barnes, Vogelsberger, Pearce, Pop, Kannan,
  Cao, Kay  \& Hernquist}{Barnes et~al.}{2021}]{Barnes2020}
Barnes D.~J.,  Vogelsberger M.,  Pearce F.~A.,  Pop A.-R.,  Kannan R.,  Cao K.,
   Kay S.~T.,   Hernquist L.,  2021, \mn@doi [Monthly Notices of the Royal
  Astronomical Society] {10.1093/mnras/stab1276}

\bibitem[\protect\citeauthoryear{{Ben{\'\i}tez-Llambay}, {Frenk}, {Ludlow}  \&
  {Navarro}}{{Ben{\'\i}tez-Llambay} et~al.}{2019}]{2019MNRAS.488.2387B}
{Ben{\'\i}tez-Llambay} A.,  {Frenk} C.~S.,  {Ludlow} A.~D.,   {Navarro} J.~F.,
  2019, \mn@doi [\mnras] {10.1093/mnras/stz1890}, \href
  {https://ui.adsabs.harvard.edu/abs/2019MNRAS.488.2387B} {488, 2387}

\bibitem[\protect\citeauthoryear{{Blumenthal}, {Faber}, {Flores}  \&
  {Primack}}{{Blumenthal} et~al.}{1986}]{blumenthal86}
{Blumenthal} G.~R.,  {Faber} S.~M.,  {Flores} R.,   {Primack} J.~R.,  1986,
  \mn@doi [\apj] {10.1086/163867}, \href
  {https://ui.adsabs.harvard.edu/abs/1986ApJ...301...27B} {301, 27}

\bibitem[\protect\citeauthoryear{{Bolton}, {Burles}, {Koopmans}, {Treu}  \&
  {Moustakas}}{{Bolton} et~al.}{2006}]{bolton06}
{Bolton} A.~S.,  {Burles} S.,  {Koopmans} L.~V.~E.,  {Treu} T.,   {Moustakas}
  L.~A.,  2006, \mn@doi [\apj] {10.1086/498884}, \href
  {http://adsabs.harvard.edu/abs/2006ApJ...638..703B} {638, 703}

\bibitem[\protect\citeauthoryear{{Brinckmann}, {Zavala}, {Rapetti}, {Hansen}
  \& {Vogelsberger}}{{Brinckmann} et~al.}{2018}]{brinckmann18}
{Brinckmann} T.,  {Zavala} J.,  {Rapetti} D.,  {Hansen} S.~H.,   {Vogelsberger}
  M.,  2018, \mn@doi [\mnras] {10.1093/mnras/stx2782}, \href
  {https://ui.adsabs.harvard.edu/abs/2018MNRAS.474..746B} {474, 746}

\bibitem[\protect\citeauthoryear{{Bryan} \& {Norman}}{{Bryan} \&
  {Norman}}{1998}]{bryan98}
{Bryan} G.~L.,  {Norman} M.~L.,  1998, \mn@doi [\apj] {10.1086/305262}, \href
  {http://adsabs.harvard.edu/abs/1998ApJ...495...80B} {495, 80}

\bibitem[\protect\citeauthoryear{{Bullock} \& {Boylan-Kolchin}}{{Bullock} \&
  {Boylan-Kolchin}}{2017}]{bullock17}
{Bullock} J.~S.,  {Boylan-Kolchin} M.,  2017, \mn@doi [\araa]
  {10.1146/annurev-astro-091916-055313}, \href
  {https://ui.adsabs.harvard.edu/abs/2017ARA&A..55..343B} {55, 343}

\bibitem[\protect\citeauthoryear{{Buote}, {Jeltema}, {Canizares}  \&
  {Garmire}}{{Buote} et~al.}{2002}]{buote02}
{Buote} D.~A.,  {Jeltema} T.~E.,  {Canizares} C.~R.,   {Garmire} G.~P.,  2002,
  \mn@doi [\apj] {10.1086/342158}, \href
  {https://ui.adsabs.harvard.edu/abs/2002ApJ...577..183B} {577, 183}

\bibitem[\protect\citeauthoryear{{Burger} \& {Zavala}}{{Burger} \&
  {Zavala}}{2021}]{burger21}
{Burger} J.~D.,  {Zavala} J.,  2021, \mn@doi [\apj] {10.3847/1538-4357/ac1a0f},
  \href {https://ui.adsabs.harvard.edu/abs/2021ApJ...921..126B} {921, 126}

\bibitem[\protect\citeauthoryear{{Chua}, {Pillepich}, {Vogelsberger}  \&
  {Hernquist}}{{Chua} et~al.}{2019}]{chua19}
{Chua} K. T.~E.,  {Pillepich} A.,  {Vogelsberger} M.,   {Hernquist} L.,  2019,
  \mn@doi [\mnras] {10.1093/mnras/sty3531}, \href
  {https://ui.adsabs.harvard.edu/abs/2019MNRAS.484..476C} {484, 476}

\bibitem[\protect\citeauthoryear{{Chua}, {Dibert}, {Vogelsberger}  \&
  {Zavala}}{{Chua} et~al.}{2021}]{chua21}
{Chua} K. T.~E.,  {Dibert} K.,  {Vogelsberger} M.,   {Zavala} J.,  2021,
  \mn@doi [\mnras] {10.1093/mnras/staa3315}, \href
  {https://ui.adsabs.harvard.edu/abs/2021MNRAS.500.1531C} {500, 1531}

\bibitem[\protect\citeauthoryear{{Col{\'{\i}}n}, {Avila-Reese}, {Valenzuela}
  \& {Firmani}}{{Col{\'{\i}}n} et~al.}{2002}]{colin02}
{Col{\'{\i}}n} P.,  {Avila-Reese} V.,  {Valenzuela} O.,   {Firmani} C.,  2002,
  \mn@doi [\apj] {10.1086/344259}, \href
  {http://adsabs.harvard.edu/abs/2002ApJ...581..777C} {581, 777}

\bibitem[\protect\citeauthoryear{{Correa}}{{Correa}}{2021}]{correa21}
{Correa} C.~A.,  2021, \mn@doi [\mnras] {10.1093/mnras/stab506}, \href
  {https://ui.adsabs.harvard.edu/abs/2021MNRAS.503..920C} {503, 920}

\bibitem[\protect\citeauthoryear{{Dav{\'e}}, {Spergel}, {Steinhardt}  \&
  {Wandelt}}{{Dav{\'e}} et~al.}{2001}]{dave01}
{Dav{\'e}} R.,  {Spergel} D.~N.,  {Steinhardt} P.~J.,   {Wandelt} B.~D.,  2001,
  \mn@doi [\apj] {10.1086/318417}, \href
  {http://adsabs.harvard.edu/abs/2001ApJ...547..574D} {547, 574}

\bibitem[\protect\citeauthoryear{Despali \& Vegetti}{Despali \&
  Vegetti}{2017}]{despali17b}
Despali G.,  Vegetti S.,  2017, \mn@doi [Monthly Notices of the Royal
  Astronomical Society] {10.1093/mnras/stx966}, 469, 1997

\bibitem[\protect\citeauthoryear{Despali, Tormen  \& Sheth}{Despali
  et~al.}{2013}]{Despali:2013}
Despali G.,  Tormen G.,   Sheth R.~K.,  2013, \mn@doi [Monthly Notices of the
  Royal Astronomical Society] {10.1093/mnras/stt235}, 431, 1143

\bibitem[\protect\citeauthoryear{{Despali}, {Giocoli}  \& {Tormen}}{{Despali}
  et~al.}{2014}]{despali14}
{Despali} G.,  {Giocoli} C.,   {Tormen} G.,  2014, \mn@doi [\mnras]
  {10.1093/mnras/stu1393}, \href
  {https://ui.adsabs.harvard.edu/abs/2014MNRAS.443.3208D} {443, 3208}

\bibitem[\protect\citeauthoryear{{Despali}, {Giocoli}, {Bonamigo}, {Limousin}
  \& {Tormen}}{{Despali} et~al.}{2017}]{despali17}
{Despali} G.,  {Giocoli} C.,  {Bonamigo} M.,  {Limousin} M.,   {Tormen} G.,
  2017, \mn@doi [\mnras] {10.1093/mnras/stw3129}, \href
  {http://adsabs.harvard.edu/abs/2017MNRAS.466..181D} {466, 181}

\bibitem[\protect\citeauthoryear{Despali, Sparre, Vegetti, Vogelsberger, Zavala
   \& Marinacci}{Despali et~al.}{2019}]{Despali:2019}
Despali G.,  Sparre M.,  Vegetti S.,  Vogelsberger M.,  Zavala J.,   Marinacci
  F.,  2019, \mn@doi [Monthly Notices of the Royal Astronomical Society]
  {10.1093/mnras/stz273}, 484, 4563

\bibitem[\protect\citeauthoryear{Eckert, Ettori, Robertson, Massey,
  Pointecouteau, Harvey  \& McCarthy}{Eckert et~al.}{2022}]{eckert22}
Eckert D.,  Ettori S.,  Robertson A.,  Massey R.,  Pointecouteau E.,  Harvey
  D.,   McCarthy I.,  2022, arXiv preprint arXiv:2205.01123

\bibitem[\protect\citeauthoryear{{Enzi}, {Vegetti}, {Despali}, {Hsueh}  \&
  {Metcalf}}{{Enzi} et~al.}{2020}]{enzi20}
{Enzi} W.,  {Vegetti} S.,  {Despali} G.,  {Hsueh} J.-W.,   {Metcalf} R.~B.,
  2020, \mn@doi [\mnras] {10.1093/mnras/staa1224}, \href
  {https://ui.adsabs.harvard.edu/abs/2020MNRAS.496.1718E} {496, 1718}

\bibitem[\protect\citeauthoryear{{Garrison-Kimmel} et~al.,}{{Garrison-Kimmel}
  et~al.}{2019}]{garrison19}
{Garrison-Kimmel} S.,  et~al., 2019, \mn@doi [\mnras] {10.1093/mnras/stz1317},
  \href {https://ui.adsabs.harvard.edu/abs/2019MNRAS.487.1380G} {487, 1380}

\bibitem[\protect\citeauthoryear{{Gavazzi}, {Treu}, {Rhodes}, {Koopmans},
  {Bolton}, {Burles}, {Massey}  \& {Moustakas}}{{Gavazzi}
  et~al.}{2007}]{gavazzi07}
{Gavazzi} R.,  {Treu} T.,  {Rhodes} J.~D.,  {Koopmans} L. V.~E.,  {Bolton}
  A.~S.,  {Burles} S.,  {Massey} R.~J.,   {Moustakas} L.~A.,  2007, \mn@doi
  [\apj] {10.1086/519237}, \href
  {https://ui.adsabs.harvard.edu/abs/2007ApJ...667..176G} {667, 176}

\bibitem[\protect\citeauthoryear{{Genel} et~al.,}{{Genel}
  et~al.}{2014}]{Genel14}
{Genel} S.,  et~al., 2014, \mn@doi [\mnras] {10.1093/mnras/stu1654}, \href
  {http://adsabs.harvard.edu/abs/2014MNRAS.445..175G} {445, 175}

\bibitem[\protect\citeauthoryear{{Gnedin}, {Kravtsov}, {Klypin}  \&
  {Nagai}}{{Gnedin} et~al.}{2004}]{gnedin04}
{Gnedin} O.~Y.,  {Kravtsov} A.~V.,  {Klypin} A.~A.,   {Nagai} D.,  2004,
  \mn@doi [\apj] {10.1086/424914}, \href
  {https://ui.adsabs.harvard.edu/abs/2004ApJ...616...16G} {616, 16}

\bibitem[\protect\citeauthoryear{{Golse} \& {Kneib}}{{Golse} \&
  {Kneib}}{2002}]{golse02}
{Golse} G.,  {Kneib} J.~P.,  2002, \mn@doi [\aap] {10.1051/0004-6361:20020639},
  \href {https://ui.adsabs.harvard.edu/abs/2002A&A...390..821G} {390, 821}

\bibitem[\protect\citeauthoryear{Harris et~al.,}{Harris et~al.}{2020}]{numpy}
Harris C.~R.,  et~al., 2020, \mn@doi [Nature] {10.1038/s41586-020-2649-2}, 585,
  357

\bibitem[\protect\citeauthoryear{{Harvey}, {Chisari}, {Robertson}  \&
  {McCarthy}}{{Harvey} et~al.}{2021}]{harvey21}
{Harvey} D.,  {Chisari} N.~E.,  {Robertson} A.,   {McCarthy} I.~G.,  2021,
  \mn@doi [\mnras] {10.1093/mnras/stab1741}, \href
  {https://ui.adsabs.harvard.edu/abs/2021MNRAS.506..441H} {506, 441}

\bibitem[\protect\citeauthoryear{Hunter}{Hunter}{2007}]{matplotlib}
Hunter J.~D.,  2007, \mn@doi [Computing In Science \& Engineering]
  {10.1109/MCSE.2007.55}, 9, 90

\bibitem[\protect\citeauthoryear{Jones, Oliphant, Peterson  et~al.}{Jones
  et~al.}{01  }]{scipy}
Jones E.,  Oliphant T.,  Peterson P.,   et~al., 2001--, {SciPy}: Open source
  scientific tools for {Python}, \url {http://www.scipy.org/}

\bibitem[\protect\citeauthoryear{{Kaplinghat}, {Tulin}  \& {Yu}}{{Kaplinghat}
  et~al.}{2014}]{kaplinghat14}
{Kaplinghat} M.,  {Tulin} S.,   {Yu} H.-B.,  2014, \mn@doi [\prd]
  {10.1103/PhysRevD.89.035009}, \href
  {https://ui.adsabs.harvard.edu/abs/2014PhRvD..89c5009K} {89, 035009}

\bibitem[\protect\citeauthoryear{{Kaplinghat}, {Valli}  \& {Yu}}{{Kaplinghat}
  et~al.}{2019}]{kaplinghat19}
{Kaplinghat} M.,  {Valli} M.,   {Yu} H.-B.,  2019, \mn@doi [\mnras]
  {10.1093/mnras/stz2511}, \href
  {https://ui.adsabs.harvard.edu/abs/2019MNRAS.490..231K} {490, 231}

\bibitem[\protect\citeauthoryear{{Kaplinghat}, {Ren}  \& {Yu}}{{Kaplinghat}
  et~al.}{2020}]{kaplinghat20}
{Kaplinghat} M.,  {Ren} T.,   {Yu} H.-B.,  2020, \mn@doi [\jcap]
  {10.1088/1475-7516/2020/06/027}, \href
  {https://ui.adsabs.harvard.edu/abs/2020JCAP...06..027K} {2020, 027}

\bibitem[\protect\citeauthoryear{{Kim}, {Peter}  \& {Hargis}}{{Kim}
  et~al.}{2018}]{kim18}
{Kim} S.~Y.,  {Peter} A. H.~G.,   {Hargis} J.~R.,  2018, \mn@doi [\prl]
  {10.1103/PhysRevLett.121.211302}, \href
  {https://ui.adsabs.harvard.edu/abs/2018PhRvL.121u1302K} {121, 211302}

\bibitem[\protect\citeauthoryear{{Lovell}, {Zavala}  \&
  {Vogelsberger}}{{Lovell} et~al.}{2019}]{lovell19}
{Lovell} M.~R.,  {Zavala} J.,   {Vogelsberger} M.,  2019, \mn@doi [\mnras]
  {10.1093/mnras/stz766}, \href
  {https://ui.adsabs.harvard.edu/abs/2019MNRAS.485.5474L} {485, 5474}

\bibitem[\protect\citeauthoryear{{Mashchenko}, {Couchman}  \&
  {Wadsley}}{{Mashchenko} et~al.}{2006}]{mashchenko06}
{Mashchenko} S.,  {Couchman} H.~M.~P.,   {Wadsley} J.,  2006, \mn@doi [\nat]
  {10.1038/nature04944}, \href
  {https://ui.adsabs.harvard.edu/abs/2006Natur.442..539M} {442, 539}

\bibitem[\protect\citeauthoryear{{McDaniel}, {Jeltema}  \&
  {Profumo}}{{McDaniel} et~al.}{2021}]{mcdaniel21}
{McDaniel} A.,  {Jeltema} T.,   {Profumo} S.,  2021, \mn@doi [\jcap]
  {10.1088/1475-7516/2021/05/020}, \href
  {https://ui.adsabs.harvard.edu/abs/2021JCAP...05..020M} {2021, 020}

\bibitem[\protect\citeauthoryear{{Miralda-Escud{\'e}}}{{Miralda-Escud{\'e}}}{2002}]{miralda02}
{Miralda-Escud{\'e}} J.,  2002, \mn@doi [\apj] {10.1086/324138}, \href
  {https://ui.adsabs.harvard.edu/abs/2002ApJ...564...60M} {564, 60}

\bibitem[\protect\citeauthoryear{{Navarro}, {Frenk}  \& {White}}{{Navarro}
  et~al.}{1997}]{navarro97}
{Navarro} J.~F.,  {Frenk} C.~S.,   {White} S.~D.~M.,  1997, \mn@doi [\apj]
  {10.1086/304888}, \href {http://adsabs.harvard.edu/abs/1997ApJ...490..493N}
  {490, 493}

\bibitem[\protect\citeauthoryear{{O{\~n}orbe}, {Boylan-Kolchin}, {Bullock},
  {Hopkins}, {Kere{\v{s}}}, {Faucher-Gigu{\`e}re}, {Quataert}  \&
  {Murray}}{{O{\~n}orbe} et~al.}{2015}]{2015MNRAS.454.2092O}
{O{\~n}orbe} J.,  {Boylan-Kolchin} M.,  {Bullock} J.~S.,  {Hopkins} P.~F.,
  {Kere{\v{s}}} D.,  {Faucher-Gigu{\`e}re} C.-A.,  {Quataert} E.,   {Murray}
  N.,  2015, \mn@doi [\mnras] {10.1093/mnras/stv2072}, \href
  {https://ui.adsabs.harvard.edu/abs/2015MNRAS.454.2092O} {454, 2092}

\bibitem[\protect\citeauthoryear{{Pakmor} \& {Springel}}{{Pakmor} \&
  {Springel}}{2013}]{2013MNRAS.432..176P}
{Pakmor} R.,  {Springel} V.,  2013, \mn@doi [\mnras] {10.1093/mnras/stt428},
  \href {https://ui.adsabs.harvard.edu/abs/2013MNRAS.432..176P} {432, 176}

\bibitem[\protect\citeauthoryear{Peter, Rocha, Bullock  \& Kaplinghat}{Peter
  et~al.}{2013}]{Peter:2013II}
Peter A. H.~G.,  Rocha M.,  Bullock J.~S.,   Kaplinghat M.,  2013, \mn@doi
  [Monthly Notices of the Royal Astronomical Society] {10.1093/mnras/sts535},
  430, 105

\bibitem[\protect\citeauthoryear{{Pillepich} et~al.,}{{Pillepich}
  et~al.}{2018}]{pillepich18}
{Pillepich} A.,  et~al., 2018, \mn@doi [\mnras] {10.1093/mnras/stx2656}, \href
  {http://adsabs.harvard.edu/abs/2018MNRAS.473.4077P} {473, 4077}

\bibitem[\protect\citeauthoryear{{Pontzen} \& {Governato}}{{Pontzen} \&
  {Governato}}{2012}]{pontzen10}
{Pontzen} A.,  {Governato} F.,  2012, \mn@doi [\mnras]
  {10.1111/j.1365-2966.2012.20571.x}, \href
  {https://ui.adsabs.harvard.edu/abs/2012MNRAS.421.3464P} {421, 3464}

\bibitem[\protect\citeauthoryear{{Read}, {Agertz}  \& {Collins}}{{Read}
  et~al.}{2016}]{2016MNRAS.459.2573R}
{Read} J.~I.,  {Agertz} O.,   {Collins} M.~L.~M.,  2016, \mn@doi [\mnras]
  {10.1093/mnras/stw713}, \href
  {https://ui.adsabs.harvard.edu/abs/2016MNRAS.459.2573R} {459, 2573}

\bibitem[\protect\citeauthoryear{{Ritondale}, {Auger}, {Vegetti}  \&
  {McKean}}{{Ritondale} et~al.}{2019}]{ritondale19a}
{Ritondale} E.,  {Auger} M.~W.,  {Vegetti} S.,   {McKean} J.~P.,  2019, \mn@doi
  [\mnras] {10.1093/mnras/sty2833}, \href
  {http://adsabs.harvard.edu/abs/2019MNRAS.482.4744R} {482, 4744}

\bibitem[\protect\citeauthoryear{{Robertson} et~al.,}{{Robertson}
  et~al.}{2018}]{robertson18}
{Robertson} A.,  et~al., 2018, \mn@doi [\mnras] {10.1093/mnrasl/sly024}, \href
  {http://adsabs.harvard.edu/abs/2018MNRAS.476L..20R} {476, L20}

\bibitem[\protect\citeauthoryear{{Robertson}, {Massey}, {Eke}, {Schaye}  \&
  {Theuns}}{{Robertson} et~al.}{2021}]{robertson21}
{Robertson} A.,  {Massey} R.,  {Eke} V.,  {Schaye} J.,   {Theuns} T.,  2021,
  \mn@doi [\mnras] {10.1093/mnras/staa3954}, \href
  {https://ui.adsabs.harvard.edu/abs/2021MNRAS.501.4610R} {501, 4610}

\bibitem[\protect\citeauthoryear{{Rocha}, {Peter}, {Bullock}, {Kaplinghat},
  {Garrison-Kimmel}, {O{\~n}orbe}  \& {Moustakas}}{{Rocha}
  et~al.}{2013}]{rocha13}
{Rocha} M.,  {Peter} A.~H.~G.,  {Bullock} J.~S.,  {Kaplinghat} M.,
  {Garrison-Kimmel} S.,  {O{\~n}orbe} J.,   {Moustakas} L.~A.,  2013, \mn@doi
  [\mnras] {10.1093/mnras/sts514}, \href
  {http://adsabs.harvard.edu/abs/2013MNRAS.430...81R} {430, 81}

\bibitem[\protect\citeauthoryear{{Roszkowski}, {Sessolo}  \&
  {Trojanowski}}{{Roszkowski} et~al.}{2018}]{2018RPPh...81f6201R}
{Roszkowski} L.,  {Sessolo} E.~M.,   {Trojanowski} S.,  2018, \mn@doi [Reports
  on Progress in Physics] {10.1088/1361-6633/aab913}, \href
  {https://ui.adsabs.harvard.edu/abs/2018RPPh...81f6201R} {81, 066201}

\bibitem[\protect\citeauthoryear{{Sameie}, {Creasey}, {Yu}, {Sales},
  {Vogelsberger}  \& {Zavala}}{{Sameie} et~al.}{2018}]{sameie18}
{Sameie} O.,  {Creasey} P.,  {Yu} H.-B.,  {Sales} L.~V.,  {Vogelsberger} M.,
  {Zavala} J.,  2018, \mn@doi [\mnras] {10.1093/mnras/sty1516}, \href
  {http://adsabs.harvard.edu/abs/2018MNRAS.479..359S} {479, 359}

\bibitem[\protect\citeauthoryear{{Sameie}, {Yu}, {Sales}, {Vogelsberger}  \&
  {Zavala}}{{Sameie} et~al.}{2020}]{sameie20}
{Sameie} O.,  {Yu} H.-B.,  {Sales} L.~V.,  {Vogelsberger} M.,   {Zavala} J.,
  2020, \mn@doi [\prl] {10.1103/PhysRevLett.124.141102}, \href
  {https://ui.adsabs.harvard.edu/abs/2020PhRvL.124n1102S} {124, 141102}

\bibitem[\protect\citeauthoryear{{Schaller} et~al.,}{{Schaller}
  et~al.}{2015}]{schaller15}
{Schaller} M.,  et~al., 2015, \mn@doi [\mnras] {10.1093/mnras/stv1067}, \href
  {http://adsabs.harvard.edu/abs/2015MNRAS.451.1247S} {451, 1247}

\bibitem[\protect\citeauthoryear{{Schaye}, {Crain}, {Bower}, {Furlong},
  {Schaller}, {Theuns}, {Dalla Vecchia}  \& {Frenk}}{{Schaye}
  et~al.}{2015}]{schaye15}
{Schaye} J.,  {Crain} R.~A.,  {Bower} R.~G.,  {Furlong} M.,  {Schaller} M.,
  {Theuns} T.,  {Dalla Vecchia} C.,   {Frenk} C.~S. e.~a.,  2015, \mn@doi
  [\mnras] {10.1093/mnras/stu2058}, \href
  {http://adsabs.harvard.edu/abs/2015MNRAS.446..521S} {446, 521}

\bibitem[\protect\citeauthoryear{{Shen}, {Brinckmann}, {Rapetti},
  {Vogelsberger}, {Mantz}, {Zavala}  \& {Allen}}{{Shen} et~al.}{2022}]{shen22}
{Shen} X.,  {Brinckmann} T.,  {Rapetti} D.,  {Vogelsberger} M.,  {Mantz} A.,
  {Zavala} J.,   {Allen} S.~W.,  2022, arXiv e-prints, \href
  {https://ui.adsabs.harvard.edu/abs/2022arXiv220200038S} {p. arXiv:2202.00038}

\bibitem[\protect\citeauthoryear{{Smith}, {Brickhouse}, {Liedahl}  \&
  {Raymond}}{{Smith} et~al.}{2001}]{smith01}
{Smith} R.~K.,  {Brickhouse} N.~S.,  {Liedahl} D.~A.,   {Raymond} J.~C.,  2001,
  \mn@doi [\apjl] {10.1086/322992}, \href
  {https://ui.adsabs.harvard.edu/abs/2001ApJ...556L..91S} {556, L91}

\bibitem[\protect\citeauthoryear{{Sonnenfeld}, {Gavazzi}, {Suyu}, {Treu}  \&
  {Marshall}}{{Sonnenfeld} et~al.}{2013}]{sonnenfeld13}
{Sonnenfeld} A.,  {Gavazzi} R.,  {Suyu} S.~H.,  {Treu} T.,   {Marshall} P.~J.,
  2013, \mn@doi [\apj] {10.1088/0004-637X/777/2/97}, \href
  {https://ui.adsabs.harvard.edu/abs/2013ApJ...777...97S} {777, 97}

\bibitem[\protect\citeauthoryear{{Sparre} \& {Springel}}{{Sparre} \&
  {Springel}}{2016}]{2016MNRAS.462.2418S}
{Sparre} M.,  {Springel} V.,  2016, \mn@doi [\mnras] {10.1093/mnras/stw1793},
  \href {http://adsabs.harvard.edu/abs/2016MNRAS.462.2418S} {462, 2418}

\bibitem[\protect\citeauthoryear{{Sparre}, {Pfrommer}  \& {Ehlert}}{{Sparre}
  et~al.}{2020}]{2020MNRAS.499.4261S}
{Sparre} M.,  {Pfrommer} C.,   {Ehlert} K.,  2020, \mn@doi [\mnras]
  {10.1093/mnras/staa3177}, \href
  {https://ui.adsabs.harvard.edu/abs/2020MNRAS.499.4261S} {499, 4261}

\bibitem[\protect\citeauthoryear{{Springel} et~al.,}{{Springel}
  et~al.}{2005}]{springel05b}
{Springel} V.,  et~al., 2005, \mn@doi [Nature] {10.1038/nature03597}, \href
  {http://adsabs.harvard.edu/abs/2005Natur.435..629S} {435, 629}

\bibitem[\protect\citeauthoryear{{Tollet} et~al.,}{{Tollet}
  et~al.}{2016}]{2016MNRAS.456.3542T}
{Tollet} E.,  et~al., 2016, \mn@doi [\mnras] {10.1093/mnras/stv2856}, \href
  {https://ui.adsabs.harvard.edu/abs/2016MNRAS.456.3542T} {456, 3542}

\bibitem[\protect\citeauthoryear{{Torrey}, {Vogelsberger}, {Genel}, {Sijacki},
  {Springel}  \& {Hernquist}}{{Torrey} et~al.}{2014}]{Torrey14}
{Torrey} P.,  {Vogelsberger} M.,  {Genel} S.,  {Sijacki} D.,  {Springel} V.,
  {Hernquist} L.,  2014, \mn@doi [\mnras] {10.1093/mnras/stt2295}, \href
  {http://adsabs.harvard.edu/abs/2014MNRAS.438.1985T} {438, 1985}

\bibitem[\protect\citeauthoryear{{Tulin} \& {Yu}}{{Tulin} \&
  {Yu}}{2018}]{tulin18}
{Tulin} S.,  {Yu} H.-B.,  2018, \mn@doi [\physrep]
  {10.1016/j.physrep.2017.11.004}, \href
  {https://ui.adsabs.harvard.edu/abs/2018PhR...730....1T} {730, 1}

\bibitem[\protect\citeauthoryear{{Velliscig} et~al.,}{{Velliscig}
  et~al.}{2015}]{velliscig15}
{Velliscig} M.,  et~al., 2015, \mn@doi [\mnras] {10.1093/mnras/stv1690}, \href
  {https://ui.adsabs.harvard.edu/abs/2015MNRAS.453..721V} {453, 721}

\bibitem[\protect\citeauthoryear{{Verde}, {Treu}  \& {Riess}}{{Verde}
  et~al.}{2019}]{verde19}
{Verde} L.,  {Treu} T.,   {Riess} A.~G.,  2019, \mn@doi [Nature Astronomy]
  {10.1038/s41550-019-0902-0}, \href
  {https://ui.adsabs.harvard.edu/abs/2019NatAs...3..891V} {3, 891}

\bibitem[\protect\citeauthoryear{{Vogelsberger}, {Zavala}  \&
  {Loeb}}{{Vogelsberger} et~al.}{2012}]{vogel12}
{Vogelsberger} M.,  {Zavala} J.,   {Loeb} A.,  2012, \mn@doi [\mnras]
  {10.1111/j.1365-2966.2012.21182.x}, \href
  {http://adsabs.harvard.edu/abs/2012MNRAS.423.3740V} {423, 3740}

\bibitem[\protect\citeauthoryear{{Vogelsberger} et~al.,}{{Vogelsberger}
  et~al.}{2014a}]{vogel14}
{Vogelsberger} M.,  et~al., 2014a, \mn@doi [\mnras] {10.1093/mnras/stu1536},
  \href {http://adsabs.harvard.edu/abs/2014MNRAS.444.1518V} {444, 1518}

\bibitem[\protect\citeauthoryear{{Vogelsberger}, {Zavala}, {Simpson}  \&
  {Jenkins}}{{Vogelsberger} et~al.}{2014b}]{vogel14b}
{Vogelsberger} M.,  {Zavala} J.,  {Simpson} C.,   {Jenkins} A.,  2014b, \mn@doi
  [\mnras] {10.1093/mnras/stu1713}, \href
  {http://adsabs.harvard.edu/abs/2014MNRAS.444.3684V} {444, 3684}

\bibitem[\protect\citeauthoryear{{Vogelsberger}, {Zavala}, {Cyr-Racine},
  {Pfrommer}, {Bringmann}  \& {Sigurdson}}{{Vogelsberger}
  et~al.}{2016}]{vogel16}
{Vogelsberger} M.,  {Zavala} J.,  {Cyr-Racine} F.-Y.,  {Pfrommer} C.,
  {Bringmann} T.,   {Sigurdson} K.,  2016, \mn@doi [\mnras]
  {10.1093/mnras/stw1076}, \href
  {http://adsabs.harvard.edu/abs/2016MNRAS.460.1399V} {460, 1399}

\bibitem[\protect\citeauthoryear{{Vogelsberger}, {Zavala}, {Schutz}  \&
  {Slatyer}}{{Vogelsberger} et~al.}{2018}]{vogel18a}
{Vogelsberger} M.,  {Zavala} J.,  {Schutz} K.,   {Slatyer} T.~R.,  2018,
  preprint, \href {http://adsabs.harvard.edu/abs/2018arXiv180503203V} {}
  (\mn@eprint {arXiv} {1805.03203})

\bibitem[\protect\citeauthoryear{{Vogelsberger}, {Zavala}, {Schutz}  \&
  {Slatyer}}{{Vogelsberger} et~al.}{2019}]{vogel19}
{Vogelsberger} M.,  {Zavala} J.,  {Schutz} K.,   {Slatyer} T.~R.,  2019,
  \mn@doi [\mnras] {10.1093/mnras/stz340}, \href
  {https://ui.adsabs.harvard.edu/abs/2019MNRAS.484.5437V} {484, 5437}

\bibitem[\protect\citeauthoryear{{Vogelsberger}, {Marinacci}, {Torrey}  \&
  {Puchwein}}{{Vogelsberger} et~al.}{2020}]{vogel20}
{Vogelsberger} M.,  {Marinacci} F.,  {Torrey} P.,   {Puchwein} E.,  2020,
  \mn@doi [Nature Reviews Physics] {10.1038/s42254-019-0127-2}, \href
  {https://ui.adsabs.harvard.edu/abs/2020NatRP...2...42V} {2, 42}

\bibitem[\protect\citeauthoryear{{Weinberger} et~al.,}{{Weinberger}
  et~al.}{2017}]{2017MNRAS.465.3291W}
{Weinberger} R.,  et~al., 2017, \mn@doi [\mnras] {10.1093/mnras/stw2944}, \href
  {http://adsabs.harvard.edu/abs/2017MNRAS.465.3291W} {465, 3291}

\bibitem[\protect\citeauthoryear{{Weinberger} et~al.,}{{Weinberger}
  et~al.}{2018}]{weinberger18}
{Weinberger} R.,  et~al., 2018, \mn@doi [\mnras] {10.1093/mnras/sty1733}, \href
  {https://ui.adsabs.harvard.edu/abs/2018MNRAS.479.4056W} {479, 4056}

\bibitem[\protect\citeauthoryear{{Zavala}, {Lovell}, {Vogelsberger}  \&
  {Burger}}{{Zavala} et~al.}{2019}]{zavala19}
{Zavala} J.,  {Lovell} M.~R.,  {Vogelsberger} M.,   {Burger} J.~D.,  2019,
  \mn@doi [\prd] {10.1103/PhysRevD.100.063007}, \href
  {https://ui.adsabs.harvard.edu/abs/2019PhRvD.100f3007Z} {100, 063007}

\bibitem[\protect\citeauthoryear{{Zolotov} et~al.,}{{Zolotov}
  et~al.}{2012}]{zolotov12}
{Zolotov} A.,  et~al., 2012, \mn@doi [\apj] {10.1088/0004-637X/761/1/71}, \href
  {https://ui.adsabs.harvard.edu/abs/2012ApJ...761...71Z} {761, 71}

\bibitem[\protect\citeauthoryear{{van de Ven}, {Mandelbaum}  \& {Keeton}}{{van
  de Ven} et~al.}{2009}]{vandeven09}
{van de Ven} G.,  {Mandelbaum} R.,   {Keeton} C.~R.,  2009, \mn@doi [\mnras]
  {10.1111/j.1365-2966.2009.15167.x}, \href
  {https://ui.adsabs.harvard.edu/abs/2009MNRAS.398..607V} {398, 607}

\makeatother
\end{thebibliography}

\appendix

\section{Additional results of the KS test} \label{sec:App1}

\begin{figure*}
\centering
    \includegraphics[width=0.95 \hsize]{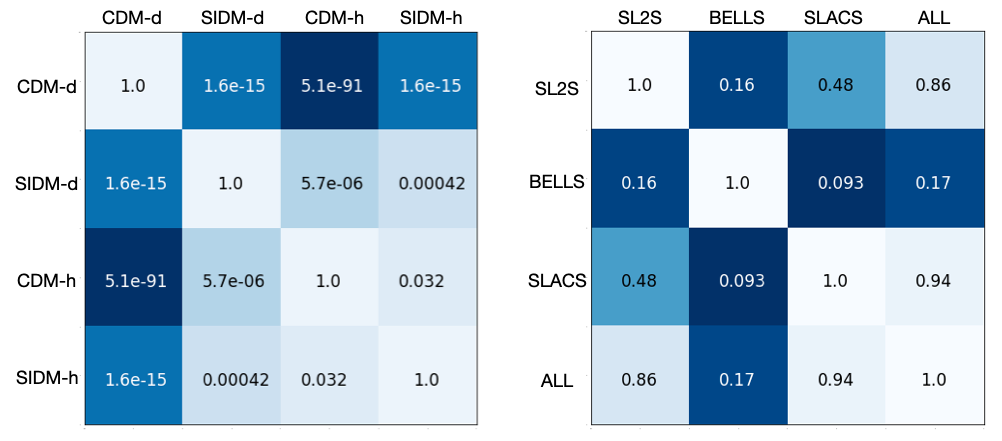}
	\caption{Results of the KS test for the simulated (left) and observed (right) samples. In each case, we show the $p$-value resulting from the comparison.
	In the cases presented here, the more compatible pairs are thus those corresponding to lighter shades of blue. In the left panel, we compare the projected ellipticity of simulated haloes at r=14 kpc, as in Figure \ref{fig:pval}.}
	\label{fig:pval2}
\end{figure*}

In Section \ref{sec:Shapes}, we compared the projected ellipticities measured from simulations to the results of gravitational lensing observations. Here we separate simulated and observed samples and discuss how they compare among each other. Figure \ref{fig:pval2} shows the $p$-value resulting from the test in each case. From the left panel, we infer that the simulated samples can be successfully distinguished in all cases: the two most similar scenarios are (as expected) the two hydro runs, but the probability of them being drawn from the same parent distribution is still only 0.032. In the case of observations, the SLACS and SL2S samples are the most compatible, while the BELLS lenses (also the smallest sample) show a departure from the general population.

In Figure \ref{fig:2Dhist} and \ref{fig:pval}, we have chosen the projected ellipticities measured within $r=14$ kpc from the centre, where this distance was chosen as an approximate estimate of the size of the region probed by lensing. However, not all observed lenses have the same angular or physical size and thus the comparison could be biased by this specific choice. 
Moreover, the observational modelling includes an external shear component that we do not have in the simulation analysis. On the other hand, here we calculate the shapes only on the projected mass distribution of the main halo (i.e. the first SUBFIND subhalo) and discard the subhalo contribution (and do not have any external source of possible shear) and thus we expect it to trace an ellipticity similar to that of the elliptical lens model.

\begin{figure*}
	\includegraphics[width=\hsize]{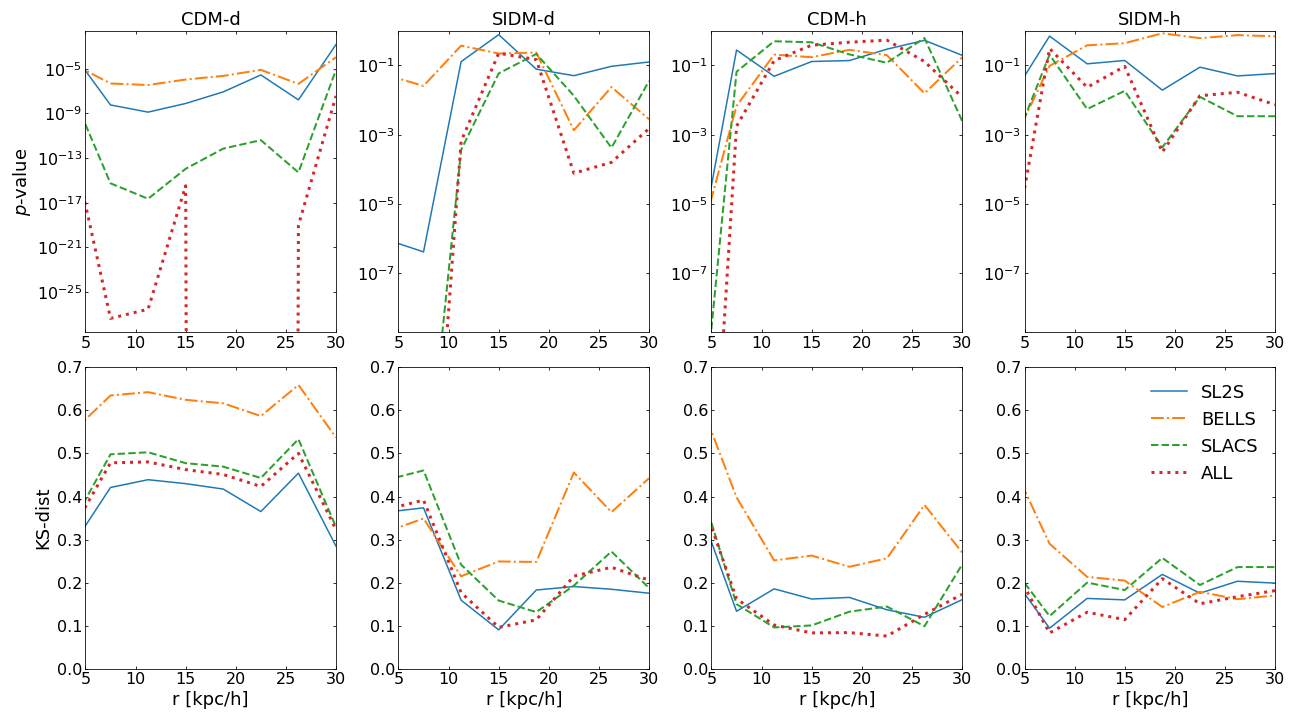}
     \caption{We show how the KS distance (bottom) and the associated $p$-value (top) of the comparison between simulations and observational data depend on radius. Lines of different colours show the results for the SL2S \citep{sonnenfeld13}, BELLS \citep{ritondale19a} and SLACS \citep{auger10b} observed ellipticity, as well as the sample including all lenses together, when compared with the simulated values (different columns). Note that the $y$-axis does not have the same limits in all panels in the top row. We remind the reader that if the KS-distance is low and the $p$-value is high, two samples are likely to be drawn from the same distribution. Conversely, if the KS-distance is high and/or the $p$-value is low (as in the leftmost panels), it is unlikely that two samples are drawn from the same parent distribution.}
    \label{fig:app1}
\end{figure*}

Here, we repeat the KS-test for different projected distances between 5 and 30 kpc$h^{-1}$ from the centre: for each radius, we compare the simulated shapes to the observational results (these do not change). The results are shown in Figure \ref{fig:app1}: in each column, we plot the $p$-value (top) and KS distance (bottom) for the comparison between one simulated data-set and the three considered lensing samples. 

It is evident that the CDM-d predictions are strongly rejected in all cases, confirming the results from Figure \ref{fig:pval}. Moreover, the SIDM-d predictions are compatible with the observational data only in the range between 10 and 20 kpc$h^{-1}$, whereas inner and outer shapes are disfavoured. Finally, the CDM-h predictions seem to provide the best agreement with observational data, followed closely by SIDM-h results. This test confirms the main conclusion drawn in Section \ref{sec:Shapes}: thanks to the addition of baryons, both cold and self-interacting dark matter are viable explanations for observed shapes.

\bsp	
\label{lastpage}
\end{document}